\documentclass[12pt]{iopart}

\usepackage{iopams, bm,graphicx,color}
\begin{document}

\title[Evaluation of Generalized Degrees of Freedom by Replica Method]
{Evaluation of Generalized Degrees of Freedom for Sparse Estimation
by Replica Method}

\author{A Sakata}

\address{The Institute of Statistical Mathematics, Midori-cho, Tachikawa 190-8562, Japan}
\address{The Graduate University for Advanced Science (SOKENDAI), Hayama-cho, Kanagawa 240-0193, Japan}
\ead{ayaka@ism.ac.jp}
\vspace{10pt}

\begin{abstract}
We develop a method to evaluate the generalized degrees of freedom (GDF)
for linear regression with sparse regularization. The GDF is a key factor in model selection, and thus its evaluation is useful in many modelling applications.
An analytical expression for the GDF is derived using
the replica method in the large-system-size limit with 
random Gaussian predictors. The resulting formula
has a universal form that is independent of the type of
regularization, providing us
with a simple interpretation.
Within the framework of replica symmetric (RS) analysis, GDF has a physical meaning as the effective fraction of
non-zero components.
The validity of our method in the RS phase is supported
by the consistency of our results with previous mathematical results.
The analytical results in the RS phase are calculated numerically using the belief propagation algorithm.
\end{abstract}
%
\noindent{\it Keywords}: Cavity and replica method, Statistical 
inference, Learning theory

%
\submitto{\JSTAT}
%
%
%

\section{Introduction}

Statistical modelling plays a key role in extracting the structures of a system that may be hidden behind observed data and using them for prediction or control.
A statistical model approximates the true generative process of the data,
which is generally expressed by a probability distribution.
Although it is necessary to adopt
an appropriate statistical model,
this will depend on the purpose of the modelling,
and the definition of appropriateness is not unique.
Akaike proposed an information criterion for model selection,
where the appropriate model is defined using Kullback--Leibler divergence \cite{AIC}.
This criterion validates the relative effectiveness of the model under consideration, and mathematically expresses the contribution of the
model to the prediction performance.

Since the systemization of the least absolute shrinkage and selection operator (LASSO) \cite{Lasso}, which simultaneously achieves variable selection and estimation,
sparse estimation has been attracting considerable attention in fields such as signal processing \cite{CS,Donoho} and machine learning \cite{ML,SVM}.
In general, sparse estimation is formulated as the problem of minimizing the estimating function penalized by sparse regularization.
The estimated variables have zero components, a property known as sparsity.
To find the sparse representation of the system from among various candidates,
a seemingly hidden rule that controls the system is sought.
Similar to LASSO, $\ell_1$ regularization is widely used
because of its convexity, which yields mathematical and algorithmic tractability \cite{CS}.
In addition, non-convex regularization, such as using the $\ell_p~(p<1)$-norm \cite{Foucart_Lai,Wang_etal},
has been studied to obtain a sparser representation than that given by $\ell_1$-norm
regularization \cite{Xu_etal}.
Furthermore, the smoothly clipped absolute deviation (SCAD) and adaptive LASSO penalty have been investigated \cite{SCAD,SCAD2,AdaptiveLasso} to acquire the oracle property, which the LASSO estimator does not possess.

The emergence of the estimation paradigm associated with sparsity
requires the development of appropriate model selection criteria.
In sparse estimation, the determination of the regularization parameter
can be regarded as the selection of a model from a family of models that have different sparsities
controlled by the regularization parameter.
In addition to the cross-validation (CV) method \cite{Stone}, which is
a simple numerical approach for sparse estimation \cite{cv_AIC,cv_obuchi},
analytical model selection methods with lower computational costs have been developed.
One such method involves estimating the {\it generalized degrees of freedom} (GDF) \cite{Ye1998}.
The GDF is a key quantity for Mallows' $C_p$, a model selection criterion based on the prediction error \cite{Mallows}.
In particular, the derivation of GDF has been studied in linear regression with a known variance \cite{Efron2004}.
The analytical form of the GDF for LASSO \cite{L1_GDF}
and elastic net regularization \cite{Zou_DThesis} are well known,
but general expressions for other regularizations have not yet been derived.

In this paper, we propose an analytical method based on statistical physics for the derivation of GDF in sparse estimation.
Certain aspects of statistical physics developed for random systems have already been applied to sparse estimation problems
\cite{CS_Kabashima,Rangan,Guo,DL_L0}.
The analysis of typical properties provides physical interpretations of 
the problems based on
phase transition pictures, and this contributes to the development of
algorithms \cite{AMP,DMM2,GAMP,Krzakala}.
The statistical physical method can be applied to the estimation of GDF for sparse regularization.
We show that GDF is expressed as the effective fraction
of non-zero components for any sparse regularization.
This expression is a mathematical realization of the meaning of GDF in terms of ``model complexity'' \cite{L1_GDF,Hastie_Tibshirani1990}.

The remainder of this paper is organized as follows.
Section \ref{sec:overview} summarizes the model selection criterion
discussed in this paper and highlights some previous related studies on sparse estimation.
Section \ref{sec:problem} explains our problem setting
for the estimation of GDF.
Sections \ref{sec:replica} and \ref{sec:replica_one_body} describe our analytical method based on the replica method for sparse estimation.
Section \ref{sec:GDF} represents the behaviour of GDF for $\ell_1$, elastic net, $\ell_0$, and SCAD regularization.
Section \ref{sec:BP} proposes the numerical calculation of GDF
using the belief propagation algorithm,
and discusses the generality of the results.
In Section \ref{sec:l0_exhaustive},
the approximation performance of our method for the calculation of GDF
is examined in the case of $\ell_0$ regularization.
Finally, Section \ref{sec:summary} concludes the paper.

\section{Overview of model selection}
\label{sec:overview}

In this section, we explain the criteria for model selection discussed in this paper.
In addition, we summarize previous studies and identify our contributions.
We focus on the parametric model, where the true generative model of $z$,
denoted by $q(z)$, is approximated by 
$p(z|\bm{\theta})$
with a parameter $\bm{\theta}\in\bm{\Theta}\subset\mathbb{R}^N$,
where $\bm{\Theta}$ is a parameter space.
The parameter is estimated under the given model to effectively describe the true distribution using training data $\bm{w}=\{w_\mu\}$ $(\mu=1,\cdots,M,w_\mu\sim q(w_\mu))$.
Let us prepare a set of candidate models ${\cal M}=\{p_1(z|\hat{\bm{\theta}}_1),\cdots,p_m(z|\hat{\bm{\theta}}_m)\}$
for the approximation of the true distribution, where $\hat{\bm{\theta}}_k$ is the estimated parameter under the $k$-th model. Model selection is then the problem of adopting a model based on a certain criterion.

\subsection{Information criterion}

The information criterion evaluates the quality of the statistical model
based on Kullback--Leibler (KL) divergence.
KL divergence describes the closeness between
the true distribution $q(z)$
and the assumed distribution $p(z|\hat{\bm{\theta}}_{\rm ML}(\bm{w}))$
as
\begin{eqnarray}
{\rm KL}(q:p)=E_{z\sim
 q(z)}
\left[\log{q(z)}\right]-
E_{z\sim q(z)}[\log{p(z|\hat{\bm{\theta}}_{\rm ML}(\bm{w}))}],
\label{eq:def_KL}
\end{eqnarray}
where $\hat{\bm{\theta}}_{\rm ML}(\bm{w})$ is the maximum likelihood estimator from the training sample $\bm{w}$.
The dependency on the model appears only in the second term of \eref{eq:def_KL}, called the predicting log-likelihood, i.e.
$l(\bm{w})\equiv E_{z\sim q(z)}[\log{p(z|\hat{\bm{\theta}}_{\rm
ML}(\bm{w}))}]$.
Therefore, the maximization of the predicting log-likelihood is
the basis for the information criterion.
Unfortunately, it is generally impossible to evaluate the predicting log-likelihood,
because we cannot determine the true distribution.
We define the estimator of the predicting log-likelihood using the empirical distribution
\begin{eqnarray}
\hat{l}(\bm{w})=\frac{1}{M}\sum_{\mu=1}^M\log 
 p(w_\mu|\hat{\bm{\theta}}_{\rm ML}(\bm{w})),
\end{eqnarray}
which corresponds to the maximum log-likelihood.
The expected value of the difference between the predicting log-likelihood
and the maximum log-likelihood, termed the bias, is given by
\begin{eqnarray}
b=E_{\bm{w}\sim q(\bm{w})}\left[\hat{l}(\bm{w})
-E_{z\sim q(z)}\left[\log p(z|\hat{\bm{\theta}}_{\rm ML}(\bm{w}))\right]\right],
\end{eqnarray}
where $q(\bm{w})=\prod_\mu q(w_\mu)$.
The information criterion is defined as
an unbiased estimator of the negative predicting log-likelihood:
\begin{eqnarray}
{\rm IC}(\bm{w})=-2\hat{l}(\bm{w})+2\hat{b}(\bm{w}),
\end{eqnarray}
where $\hat{b}(\bm{w})$ is an unbiased estimator of the bias,
and
the coefficient $2$ is a conventional value.
The optimal model is defined as that which minimizes ${\rm IC}(\bm{w})$
among the models in ${\cal M}$.
Intuitively, the first and second terms represent the
training error and the complexity of the model, respectively.
As the complexity of the model increases, the model can express various distributions.
However, overfitting is likely to occur, which hampers the prediction of unknown data.
The information criterion selects the model that achieves the best 
trade-off between the training error and 
the level of model complexity.

The values of $b$ can be calculated asymptotically.
In particular, when the statistical model
contains the true model, namely a parameter $\bm{\theta}^*$
exists such that $q(z)=p(z|\bm{\theta}^*)$,
the information criterion
is known as Akaike's information criterion (AIC),
where the bias term $b$ is reduced to the dimension of the parameter
$\bm{\theta}$ \cite{AIC}.

The criterion explained thus far is for models constructed by maximum likelihood estimation. To determine the parameter with other learning strategies, we focus on
maximum likelihood estimation under regularization, where the GDF facilitates the extension of the information criterion \cite{Hastie_Tibshirani1990}.
A general expression of GDF is naturally derived from
another model selection criterion, namely, Mallows' $C_p$ \cite{Mallows}.

%

\subsection{Mallows' $C_p$ and generalized degrees of freedom}

The prediction of unknown data is another criterion for the evaluation of a model.
We define the squared prediction error per component as
\begin{eqnarray}
{\rm err}_{\rm pre}(\bm{w})=\frac{1}{M}E_{\bm{z}}[||\bm{z}-\hat{\bm{w}}(\bm{w})||_2^2],
\end{eqnarray}
where $\hat{\bm{w}}$ is the estimate of $\bm{w}$ and 
$\bm{z}\in\mathbb{R}^{M}$ is independent of $\bm{w}\in\mathbb{R}^M$,
but each component of $\bm{z}$ is generated according to the same distribution as $\bm{w}$.
When the training sample is generated as $\bm{w}\sim{\cal N}(\bm{\mu},\sigma^2\bm{I}_M)$,
where $\bm{I}_M$ is the $M$-dimensional identity matrix,
Mallows' $C_p$, calculated as
\begin{eqnarray}
{c_p}(\bm{w})={\rm err}_{\rm
 train}(\bm{w})+2\sigma^2\hat{\rm df}(\bm{w}),
\label{eq:Cp_def}
\end{eqnarray}
is an unbiased estimator of the prediction error.
Here,
\begin{eqnarray}
{\rm err}_{\rm train}(\bm{w})&=\frac{1}{M}||\bm{w}-\hat{\bm{w}}(\bm{w})||_2^2
\end{eqnarray}
is the training error and $\hat{\rm df}(\bm{w})$
is an unbiased estimator of GDF
defined by
\begin{eqnarray}
{\rm df}=\frac{{\rm
 cov}(\bm{w},\hat{\bm{w}}(\bm{w}))}{M\sigma^2},
\label{eq:def_df}
\end{eqnarray}
which quantifies
the complexity of the model \cite{L1_GDF,Hastie_Tibshirani1990},
where ${\rm cov}(\bm{w},\hat{\bm{w}}(\bm{w}))
=E_{\bm{w}}[(\bm{w}-E_{\bm{w}}[\bm{w}])(\hat{\bm{w}}(\bm{w})-E_{\bm{w}}[\hat{\bm{w}}(\bm{w})])]$.
In the framework of $C_p$,
the optimal model is defined as that which minimizes 
$c_p(\bm{w})$ among the models in ${\cal M}$.

Another expression of GDF
is given by \cite{Ye1998}
\begin{eqnarray}
{\rm df}=\frac{1}{M}\sum_\mu E_{\bm{\bm{w}}}\left[\frac{\partial \hat{w}_\mu(\bm{w})}{\partial w_\mu}\right],
\label{eq:GDF_Ye}
\end{eqnarray}
which corresponds to the expectation of
Stein's unbiased risk estimate (SURE) for the
prediction error \cite{SURE}.  
GDF was originally introduced
as an extension of the degrees of freedom
in the linear estimation rule
for a general modelling procedure
in the form \eref{eq:GDF_Ye} \cite{Ye1998}.

When the assumed model obeys a Gaussian distribution $p(\bm{w}|\bm{\theta})\propto
\exp(-\frac{1}{2\sigma^2}||\bm{w}-\bm{\mu}(\bm{\theta})||_2^2)$
with a known variance,
and taking $\bm{\mu}(\hat{\bm{\theta}}_{\rm ML}(\bm{w}))=\hat{\bm{w}}(\bm{w})$,
AIC (normalized by the number of training samples)
is given by \cite{L1_GDF}
\begin{eqnarray}
{\rm AIC}(\bm{w})=\frac{{\rm err}_{\rm train}(\bm{w})}{\sigma^2}+2\hat{\rm df}(\bm{w}).
\label{eq:AIC_df}
\end{eqnarray}
Equations \eref{eq:Cp_def} and \eref{eq:AIC_df}
indicate that model selection based on AIC and that based on $C_p$
give the same result;
they are proportional to each other
$c_p(\bm{w})=\sigma^2{\rm AIC}(\bm{w})$.

\subsection{Model selection for sparse regularization and our contributions}

The regression problems with sparse regularization
is formulad as
\begin{eqnarray}
\min_{\bm{x}}\left\{e(\bm{x};\bm{w},\bm{A})+r(\bm{x};\eta)\right\},
\end{eqnarray}
where $e(\bm{x};\bm{w},\bm{A})$ measures the difference between
training data $\bm{w}$ and its fit using regression coefficients $\bm{x}$
under the predictor matrix $\bm{A}$,
and $r(\bm{x};\eta)$ is the regularization term
with the regularization parameter $\eta$
that enhances zero components in $\bm{x}$.
The regularization parameter determines
the number of predictors used in the expression of the data distribution,
and the model distribution under the determined number of predictors can be
regarded as a model:
${\cal M}=\{p_{\eta}(z|\hat{\bm{\theta}}_{\eta})|\eta\in\bm{H}\}$,
where $\bm{H}$ is the support of the regularization parameter.
Therefore,
tuning the regularization parameter $\eta$ 
corresponds to model selection.
However, in general, the derivation of AIC based on the asymptotic expansion
is not straightforwardly applicable to sparse regularization.
In such cases, $C_p$ is useful for deriving the model selection criterion
when the squared error is considered.
In LASSO, it is mathematically proven that, when the number of training samples is greater than
the number of predictors,
the ratio of the number of non-zero regression coefficients to
the number of training samples is
an unbiased estimator of the degrees of freedom
in a finite sample \cite{L1_GDF}.
However, the derivation of GDF is analytically difficult for general
sparse regularizations.
To overcome this difficulty, GDF computation techniques have been developed using the parametric bootstrap method \cite{Efron2004} and SURE \cite{SURE,Donoho_Johnstone}.

In the present paper, we propose
an estimation technique for GDF using the replica method
under a replica symmetric (RS) assumption for linear regression
with Gaussian i.i.d. predictors.
The replica symmetric analysis for the estimation problems
under sparse regularization
are shown in \cite{CS_Kabashima,Rangan,Guo,DL_L0}.
In these papers, the replica method is employed
to study phase transition or the property of estimators.
We extend this analytical method for the calculation of GDF
that is not taken into account in the current formalism of the
replica analysis.
The technique we propose is applicable to general sparse regularization.
Using our method,
the correspondence between GDF and 
the effective fraction of non-zero components
in the large-system-size limit
is shown to be independent of the form of regularization.
Our approach differs from previous methods in which
GDF has been derived for
specific types of regularization.
We apply our method to $\ell_1$, elastic net,
$\ell_0$, and SCAD regularization
to obtain the GDF.
The results shown here for $\ell_1$ and elastic net regularization
are weaker than those in previous studies,
where the unbiased estimator of GDF, $\hat{\rm df}$,
is derived for one instance of the predictor.
However, our method is consistent with previous results,
which supports the validity of our approach.
Furthermore, our method can be applied to 
non-convex sparse regularizations such as
$\ell_0$ and SCAD,
and extends the discussion of GDF
to general sparse regularization.
For the $\ell_0$ case, the solution under the RS assumption is always unstable
against perturbations that break the replica symmetry,
but we show that GDF under the RS assumption approximates the true value
of GDF.
In the case of SCAD regularization,
our method can identify the most appropriate model
based on the prediction error
within the range of the RS assumption when the mean of the data is sufficiently small.
The generality of the result in terms of the correspondence between
GDF and the effective fraction of non-zero components
is discussed using
a belief propagation algorithm
for other predictor matrices.

\section{Problem setting and formulation}
\label{sec:problem}

We apply a linear regression model with sparse regularization $r(\bm{x};\eta)=\sum_ir(x_i;\eta)$,
where $\eta$ is a regularization parameter,
to a set of training data $\bm{y}\in\mathbb{R}^M$:
\begin{eqnarray}
\min_{\bm{x}}\left\{\frac{1}{2}||\bm{y}-\bm{Ax}||_2^2+r(\bm{x};\eta)\right\},
\label{eq:lasso}
\end{eqnarray}
where the column vectors of $\bm{A}=\{\bm{A}_1,\cdots,\bm{A}_N\}\in\mathbb{R}^{M\times N}$
and components of $\bm{x}\in\mathbb{R}^N$
correspond to predictors and regression coefficients, respectively.
Here, the coefficient of the squared error, $1\slash 2$,
is introduced for mathematical convenience.
The variable $\bm{x}$ to be estimated here corresponds to
the parameter
$\bm{\theta}$ in the previous section,
and the number of non-zero components in $\bm{x}$
corresponds to the number of parameters used in the model.
We introduce the
posterior distribution of $\bm{x}$:
\begin{eqnarray}
P_{\beta}(\bm{x}|\bm{y},\bm{A})=\exp\left\{-\frac{\beta}{2}||\bm{y}-\bm{A
 x}||_2^2-\beta r(\bm{x};\eta)-\ln Z_\beta(\bm{y},\bm{A})\right\},
\label{eq:prob}
\end{eqnarray}
where $Z_\beta(\bm{y},\bm{A})$ is the normalization constant.
The distribution as $\beta\to\infty$
is the uniform distribution over the minimizers of \eref{eq:lasso}.
Estimate of the solution of \eref{eq:lasso} under a fixed set of $\{\bm{y},\bm{A}\}$,
denoted by $\hat{\bm{x}}(\bm{y},\bm{A})$, is given by 
\begin{eqnarray}
\hat{\bm{x}}(\bm{y},\bm{A})=\lim_{\beta\to\infty}\langle\bm{x}\rangle_{\beta},
\end{eqnarray}
where $\langle\cdot\rangle_\beta$
denotes the expectation according to \eref{eq:prob}
at $\beta$.
Using this estimate
$\hat{\bm{x}}(\bm{y},\bm{A})$
of $\bm{x}$, the training sample $\bm{y}$
is estimated as
\begin{eqnarray}
\hat{\bm{y}}(\bm{y},\bm{A})=\bm{A}\hat{\bm{x}}(\bm{y},\bm{A}).
\end{eqnarray}
To understand the typical performance of \eref{eq:lasso},
we calculate the expectation of
the training error with respect to
$\bm{y}$ and $\bm{A}$,
\begin{eqnarray}
\overline{\rm err}_{\rm train}=E_{\bm{y},\bm{A}}[{\rm err}_{\rm
 train}(\bm{y},\bm{A})].
\end{eqnarray}
At a sufficiently large system size $N\to\infty$,
we set the scaling relationship as $\alpha=M\slash N\sim O(1)$
and $\hat{\rho}=K\slash N\sim O(1)$,
where $K$ is the number of non-zero components of $\hat{\bm{x}}$.
The training error relates to the free energy density $f$ as \cite{Nakanishi}
\begin{eqnarray}
f&\equiv-\lim_{\beta\to\infty}\frac{1}{N}\frac{\partial}{\partial\beta}\phi_\beta=\frac{\alpha\overline{\rm err}_{\rm train}}{2}+\overline{r},
\end{eqnarray}
where $\phi_\beta\equiv E_{\bm{y},\bm{A}}[\ln Z_{\beta}(\bm{y},\bm{A})]$
and
\begin{eqnarray}
\overline{r}=\frac{1}{N}E_{\bm{y},\bm{A}}[r(\hat{\bm{x}}(\bm{y},\bm{A});\eta)].
\end{eqnarray}
The expectation of the regularization term $\overline{r}$
is derived separately from $f$, as shown in the following section.
Hence, the training error is derived as
\begin{eqnarray}
\overline{\rm err}_{\rm train}&=\frac{2(f-\overline{r})}{\alpha}.
\end{eqnarray}

For the calculation of GDF, we introduce external fields $\kappa$ and $\nu$,
and define the extended posterior distribution as
\begin{eqnarray}
\nonumber
P_{\beta,\kappa,\nu}(\bm{x}|\bm{y},\bm{A})&\!=\!\exp\left\{-\frac{\beta}{2}||\bm{y}\!-\!\bm{A
 x}||_2^2\!-\!\beta r(\bm{x};\eta)\!-\!\beta\!\sum_{\mu i}(\kappa y_\mu\!+\!\nu) A_{\mu i}x_i\!-\!\ln Z_{\beta,\kappa,\nu}(\bm{y},\!\bm{A})\right\},\\
\label{eq:prob_h}
\end{eqnarray}
where $Z_{\beta,\kappa,\nu}(\bm{y},\bm{A})$ is the normalization constant.
We define the extended free energy density as
\begin{eqnarray}
f_{\kappa,\nu}&=-\lim_{\beta\to\infty}\frac{1}{N}\frac{\partial}{\partial\beta}\phi_{\beta,\kappa,\nu},
\end{eqnarray}
where $\phi_{\beta,\kappa,\nu}=E_{\bm{y},\bm{A}}[\ln
Z_{\beta,\kappa,\nu}(\bm{A},\bm{y})]$ and $f=f_{\kappa=0,\nu=0}$.
We derive the following quantities from the extended free energy density:
\begin{eqnarray}
\hat{\gamma}&\equiv\frac{1}{M}\sum_\mu E_{\bm{y},\bm{A}}[y_\mu\sum_iA_{\mu i}\hat{x}_i(\bm{y},\bm{A})]=\frac{1}{\alpha}\frac{\partial}
 {\partial \kappa}f_{\kappa,\nu}\Big|_{\kappa,\nu=0}\label{eq:GDF_tmp1}\\
\hat{m}_y&\equiv\frac{1}{M}\sum_\mu E_{\bm{y},\bm{A}}[\sum_i
A_{\mu
i}\hat{x}_i(\bm{y},\bm{A})]=\frac{1}{\alpha}\frac{\partial}{\partial \nu}f_{\kappa,\nu}\Big|_{\kappa,\nu=0}.
\label{eq:GDF_tmp2}
\end{eqnarray}
Using these, the GDF for a Gaussian training sample is derived as
\begin{eqnarray}
{\rm df}=\frac{\hat{\gamma}-\hat{m}_ym_y}{\sigma_y^2},
\label{eq:df_gen}
\end{eqnarray}
where
$m_y\in\mathbb{R}$ and $\sigma_y^2\in\mathbb{R}$ are
the mean and variance of the training sample, respectively.
Further, $C_p$ given by \eref{eq:Cp_def}
is the unbiased estimator of the prediction error.
Hence, the expectation of the prediction error
with respect to $\bm{y}$ and $\bm{A}$,
\begin{eqnarray}
\overline{\rm err}_{\rm pre}=\frac{1}{M}E_{\bm{y},\bm{A}}[E_{\bm{z}}[||\bm{z}-\hat{\bm{y}}(\bm{y},\bm{A})||_2^2]],
\end{eqnarray}
is given by
\begin{eqnarray}
\overline{\rm err}_{\rm pre}=\overline{\rm err}_{\rm train}+2\sigma_y^2{\rm df}.
\label{eq:pre_error}
\end{eqnarray}

\section{Analysis}
\label{sec:replica}

For the derivation of GDF, we resort to the replica method \cite{beyond,Nishimori}.
The RS calculations for $\ell_0$
and $\ell_1$ minimization are shown in the typical performance analysis of compressed sensing \cite{CS_Kabashima} and dictionary learning \cite{DL_L0}.
We summarize the analytical method and explain how it can be extended to the evaluation of GDF. Hereafter, we consider
Gaussian i.i.d. predictors
$A_{\mu i}\sim{\cal N}(0,M^{-1})~\forall (\mu,i)$.

\subsection{Replica method and replica symmetry}

We calculate the generating function $\phi_\beta$ using the following identity:
\begin{eqnarray}
E_{\bm{y},\bm{A}}[\ln Z_{\beta}(\bm{y},\bm{A})]=
\lim_{n\to 0}\frac{E_{\bm{y},\bm{A}}[Z_{\beta}^n(\bm{y},\bm{A})]-1}{n}.
\label{eq:replica}
\end{eqnarray}
Assuming that $n$ is a positive integer,
we can express the expectation of $Z_\beta^n(\bm{y},\bm{A})$
by introducing $n$ replicated systems:
\begin{eqnarray}
\nonumber
E_{\bm{y},\bm{A}}[Z_{\beta}^n(\bm{y},\bm{A})]&=\int d\bm{A}d\bm{y}P_{A}(\bm{A})P_y(\bm{y})\int
 d\bm{x}^{(1)}\cdots d\bm{x}^{(n)}\\
&\times\exp\Big[\sum_{a=1}^n\Big\{-\frac{\beta}{2}||\bm{y}-\bm{Ax}^{(a)}
 ||_2^2-\beta r(\bm{x}^{(a)};\eta)\Big\}\Big],
\label{eq:E_replica}
\end{eqnarray}
where 
$P_A(\bm{A})=\prod_{\mu,i}\sqrt{\frac{M}{2\pi}}\exp(-\frac{M}{2}A_{\mu i}^2)$
and $P_y(\bm{y})=\prod_\mu\sqrt{\frac{1}{2\pi\sigma_y^2}}\exp(-\frac{1}{2\sigma_y^2}(y-m_y)^2)$.
We characterize the microscopic states of $\{\bm{x}^{(a)}\}$
with the macroscopic quantities
\begin{eqnarray}
q^{(ab)}&=\frac{1}{M}\sum_ix_i^{(a)}x_i^{(b)}.
\end{eqnarray}
Introducing the identity for all combinations of $a,b~(a\leq b)$
\begin{eqnarray}
1=\int dq^{(ab)}
\delta\left(q^{(ab)}-\frac{1}{M}\sum_ix_i^{(a)}x_i^{(b)}\right),
\label{eq:subshell}
\end{eqnarray}
the integration with respect to $\bm{A}$
leads to the following expression:
\begin{eqnarray}
\nonumber
E_{\bm{y},\bm{A}}[Z_{\beta}^n(\bm{y},\bm{A})]&=\int d{\cal Q}{\cal
 S}({\cal Q})\int \{d\bm{u}^{(a)}\}P_u(\{\bm{u}^{(a)}\}|{\cal Q})\int d\bm{y}P_y(\bm{y})\\
&\times\exp\Big\{-\frac{\beta}{2}\sum_a||\bm{y}-\bm{u}^{(a)}||^2_2\Big\},
\end{eqnarray}
where each component of $\bm{u}^{(a)}$,
denoted by $u_\mu^{(a)}$,
is statistically equivalent to $\sum_iA_{\mu i}x_i^{(a)}$,
and ${\cal Q}$ is a matrix representation of $\{q^{(ab)}\}$.
Setting $\tilde{\bm{u}}_\mu=\{u_\mu^{(1)},\cdots,u_\mu^{(n)}\}$,
its probability distribution 
is given by \cite{CS_Kabashima}
\begin{eqnarray}
P_u(\{\bm{u}^a\}|{\cal Q})=\prod_\mu\frac{1}{\sqrt{(2\pi)^n|{\cal
 Q|}}}\exp\Big(-\frac{1}{2}\tilde{\bm{u}}_\mu^{\rm T}{\cal Q}^{-1}\tilde{\bm{u}}_\mu\Big),
\end{eqnarray}
and the function ${\cal S}({\cal Q})$ is given by
\begin{eqnarray}
\nonumber
{\cal S}({\cal Q})=\int d\hat{\cal Q}\{d\bm{x}^{(a)}\}\exp\Big\{-M&{\sum_{a\leq b}q^{(ab)}\hat{q}^{(ab)}}+\sum_{a\leq
  b}\sum_i\hat{q}^{(ab)}x_i^{(a)}x_i^{(b)}\\
&-\beta \sum_ar(\bm{x}^{(a)};\eta)\Big\},
\end{eqnarray}
where $\hat{q}^{(ab)}$ is the conjugate variable for the integral 
representation of the delta function in \eref{eq:subshell},
and $\hat{\cal Q}$ is the matrix representation of $\{\hat{q}^{(ab)}\}$.

To obtain an analytic expression with respect to $n\in\mathbb{R}$
and take the limit as $n\to 0$,
we restrict the candidates for the dominant saddle point
to those of RS form as
\begin{eqnarray}
(q^{(ab)},\hat{q}^{(ab)})=\left\{
\begin{array}{ll}
(Q,-\tilde{Q}\slash 2) & (a=b) \\
(q,\tilde{q}) & (a\neq b).
\end{array}\right.
\end{eqnarray}
For $\beta\to\infty$, RS order parameters
scale to keep $\beta(Q-q)=\chi$,
$\beta^{-1}(\tilde{Q}+\tilde{q})=\hat{Q}$,
and $\beta^{-2}\tilde{q}=\hat{\chi}$
of the order of unity.
Under the RS assumption,
the free energy density is given by
\begin{eqnarray}
f=\mathop{\rm extr}_{Q,\chi,\hat{Q},\hat{\chi}}\Big\{\frac{\alpha(Q+\sigma_y^2+m_y^2)}{2(1+\chi)}-\frac{\alpha(Q\hat{Q}-\chi\hat{\chi})}{2}
-\frac{1}{2}\pi_r(\hat{Q},\hat{\chi})\Big\},
\end{eqnarray}
where
${\rm extr}_{Q,\chi,\hat{Q},\hat{\chi}}$
denotes extremization with respect to the
variables $\{Q,\chi,\hat{Q},\hat{\chi}\}$.
The function $\pi_r$,
where the subscript $r$
denotes the dependency on the regularization, is given by
\begin{eqnarray}
&\pi_r(\hat{Q},\hat{\chi})=2\int Dz\log g_r(h^{\rm RS}(z;\hat{\chi}),\hat{Q})\\
&g_r(h,\hat{Q})=\max_x\exp\Big(-\frac{\hat{Q}}{2}x^2+hx-r(x;\eta)\Big),\label{eq:one_body}
\label{eq:f_pi_def}
\end{eqnarray}
where $h^{\rm RS}(z;\hat{\chi})=\sqrt{\hat{\chi}}z$
is the random field that effectively represents the
randomness of the problem introduced by $\bm{y}$
and $\bm{A}$,
and $Dz=dz\exp(-z^2\slash 2)\slash\sqrt{2\pi}$.
The solution of $x$ concerned with the effective single-body problem \eref{eq:one_body},
denoted by $x^*_r(z;\hat{Q},\hat{\chi})$,
is statistically equivalent to the solution of the
original problem \eref{eq:lasso}.
Therefore,
the expectation of the regularization term is
derived as
\begin{eqnarray}
\overline{r}=\int Dz r(x^*_r(z;\hat{Q},\hat{\chi});\eta).
\end{eqnarray}

The variables $Q,\chi,\hat{Q},\hat{\chi}$
are determined
by saddle point equations to satisfy the
extremum conditions of the free energy density:
\begin{eqnarray}
\chi&=\frac{1}{\alpha}\frac{\partial\pi_r(\hat{Q},\hat{\chi})}{\partial\hat{\chi}}\label{eq:RS_chi_gen}\\
Q&=-\frac{1}{\alpha}\frac{\partial\pi_r(\hat{Q},\hat{\chi})}{\partial\hat{Q}}\label{eq:RS_Q_gen}\\
\hat{\chi}&=\frac{Q+\sigma_y^2+m_y^2}{(1+\chi)^2}\label{eq:RS_chih}\\
\hat{Q}&=\frac{1}{1+\chi}.\label{eq:RS_Qh}
\end{eqnarray}
Note that the functional form of the parameters $\hat{\chi}$
and $\hat{Q}$ does not depend on the regularization,
but the values of $\chi$ and $Q$ are regularization-dependent.
At the extremum,
the parameters $Q$ and $\chi$
are related to the physical quantities by
\begin{eqnarray}
Q&=\frac{1}{M}\sum_{i=1}^NE_{\bm{y},\bm{A}}[||\hat{\bm{x}}(\bm{y},\bm{A})||_2^2]\\
\chi&=\lim_{\beta\to\infty}\frac{\beta}{M}\sum_{i=1}^NE_{\bm{y},\bm{A}}[\langle||\bm{x}||^2_2\rangle_\beta-||\langle\bm{x}\rangle_\beta||^2_2],
\label{eq:chi_physical}
\end{eqnarray}
and can be expressed using $x^*_r$ as
\begin{eqnarray}
\chi&=\frac{1}{\alpha}\int Dz\frac{\partial x^*_r(z;\hat{Q},\hat{\chi})}{\partial(\sqrt{\hat{\chi}}z)}
\label{eq:chi_single_body}\\
Q&=\frac{1}{\alpha}\int Dz(x_r^*(z;\hat{Q},\hat{\chi}))^2.
\label{eq:Q_single_body}
\end{eqnarray}

The extended free energy density with the external fields $\kappa$ and $\nu$
is given by
\begin{eqnarray}
f_{\kappa,\nu}&=f-\left\{\frac{\alpha(m_y^2+\sigma_y^2)\kappa({\kappa}-2)}{2(1+\chi)}+\frac{\alpha\nu(\nu-2m_y)}{2(1+\chi)}\right\}\chi.
\end{eqnarray}
To evaluate $f_{\kappa,\nu}$ for non-zero $\kappa$ and $\nu$, one
has to solve the saddle point equation at non-zero
$\kappa$ and $\nu$ to determine the saddle point
value of $\chi$. However,
since one would only need to evaluate derivatives of
$f_{\kappa,\nu}$ at $\kappa=\nu=0$ to obtain GDF,
the saddle point value of $\chi$ that is to be used in such
evaluations should remain the same as that obtained in the calculation
of $f$.
From \eref{eq:GDF_tmp1}--\eref{eq:df_gen}, GDF is obtained as
\begin{eqnarray}
{\rm df}=\frac{\chi}{1+\chi}=\frac{\chi}{\hat{Q}^{-1}},
\label{eq:GDF}
\end{eqnarray}
where $\chi$ and $\hat{Q}$ satisfy
the saddle point equations \eref{eq:RS_chi_gen} and
\eref{eq:RS_Qh}, respectively.
This expression is also independent of the form of the regularization.
The effective single-body problem \eref{eq:one_body}
can be interpreted as a scalar estimation problem
in which $x$ is estimated on the basis of
the prior (regularization) $\exp(-r(x;\eta))$ and the random observation
$h\slash\hat{Q}$, which is assumed to
be generated as $h\slash\hat{Q}=x+n$,
where $n\sim{\cal N}(0,\hat{Q}^{-1})$ is the 
Gaussian observation noise.
If one uses the observation itself in
the single-body problem as an estimate of $x$,
then it is an unbiased estimator of $x$ and its variance is
$\hat{Q}^{-1}$.
However, the actual variance of the estimates can change
according to the
regularization.
The variable $\chi$ is the rescaled variance of the system
expressed as \eref{eq:chi_physical}.
Therefore, GDF \eref{eq:GDF} corresponds to the effective
fraction of the non-zero components
of $\hat{\bm{x}}$
(parameters),
which is estimated by
dividing the variance of the total system by that of one 
component
when the observation is used as the estimate.
The effective fraction of the non-zero components
is measured under the assumption that the regularization does not change
the variance of one component from $\hat{Q}^{-1}$.
If this assumption is correct 
and the fluctuation of non-zero components is the unique source of the
system's fluctuation,
GDF is considered to be equal to the ratio of the number of non-zero components
to the number of training samples.

The RS solution discussed thus far loses local stability under perturbations that break the symmetry between replicas in a certain parameter region.
Known as the de Almeida--Thouless (AT) instability \cite{AT}, this phenomenon appears when 
\begin{eqnarray}
\frac{1}{\alpha(1+\chi)^2}\int Dz\Big\{\frac{\partial x^*_r(z;\hat{Q},\hat{\chi})}{\partial(\sqrt{\hat{\chi}}z)}\Big\}^2>1.
\end{eqnarray}
In general, when AT instability appears, we have to construct the full-step replica symmetry breaking (RSB) 
solution for an exact evaluation. However, the RS solution remains meaningful as an approximation \cite{beyond,Nishimori}.

\section{Applications to several sparse regularizations}
\label{sec:replica_one_body}

As shown in the previous section,
some regularization-dependency appears in the
effective single-body problem \eref{eq:one_body}.
We now apply the analytical method to
$\ell_1$, elastic net, $\ell_0$, and SCAD regularization.
The ratio of the number of non-zero components to the number of training samples is denoted by $\delta=\hat{\rho}\slash\alpha$,
and we focus on the physical region $\delta\leq 1$,
where the number of unknown variables is smaller than the number of known variables.

\begin{figure}
\begin{center}
\includegraphics[width=3in]{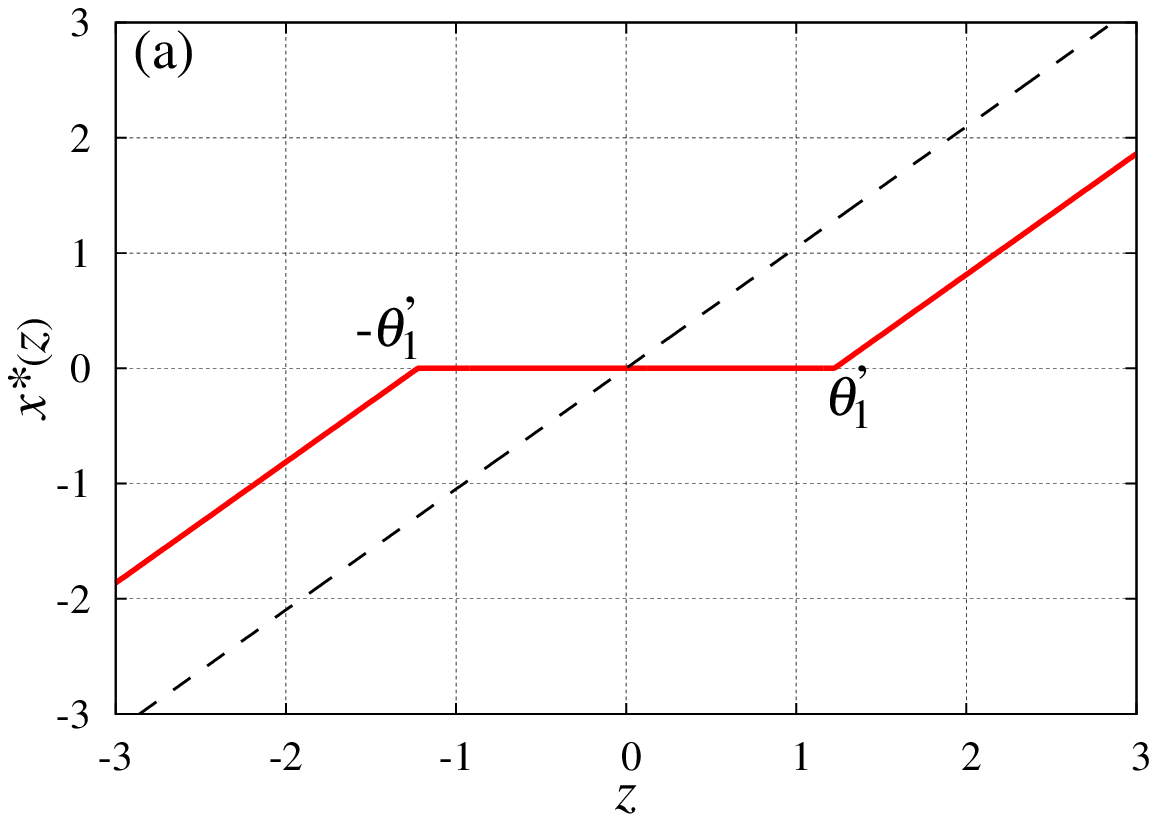}
\includegraphics[width=3in]{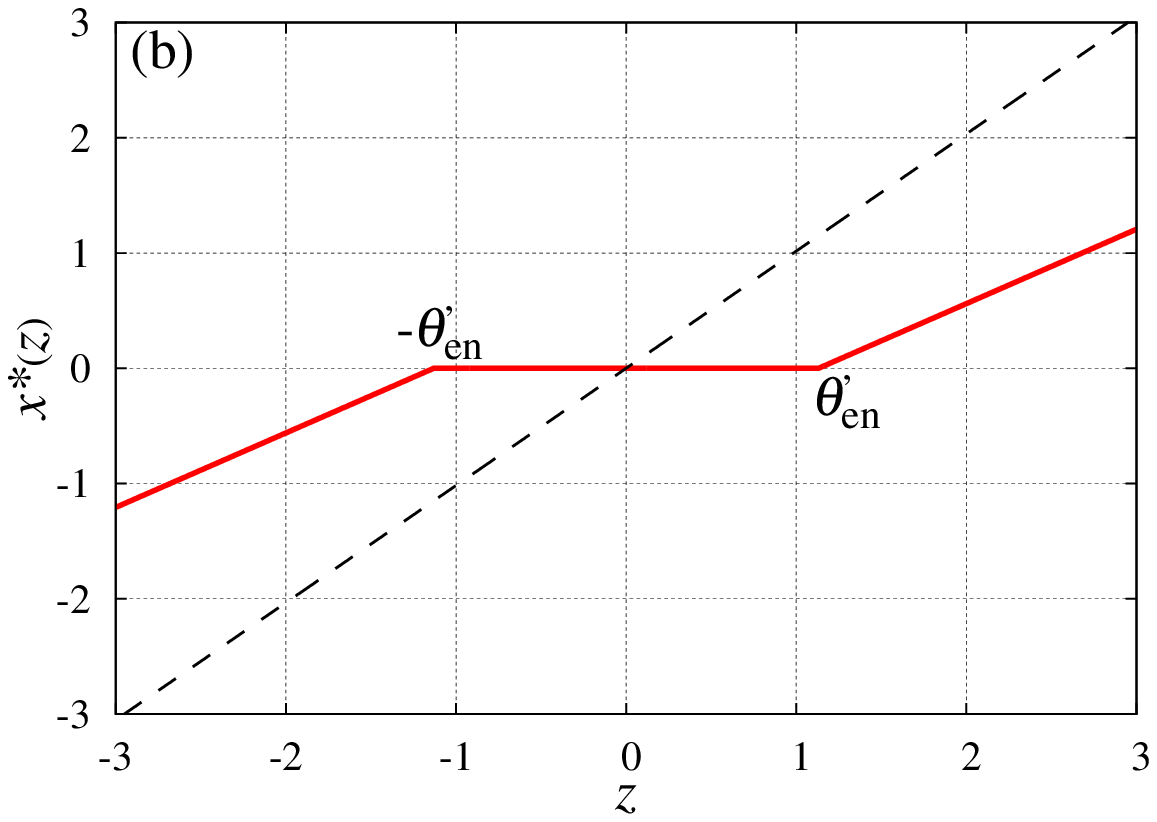}
\includegraphics[width=3in]{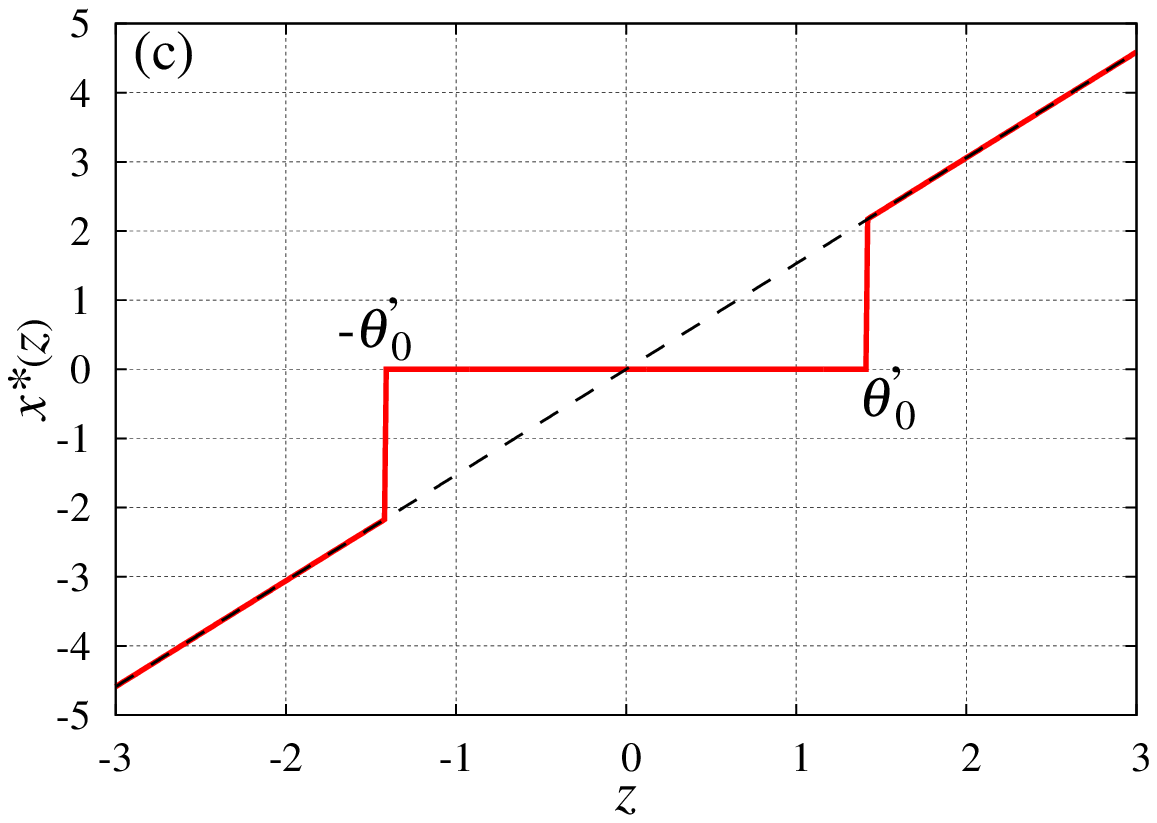}
\includegraphics[width=3in]{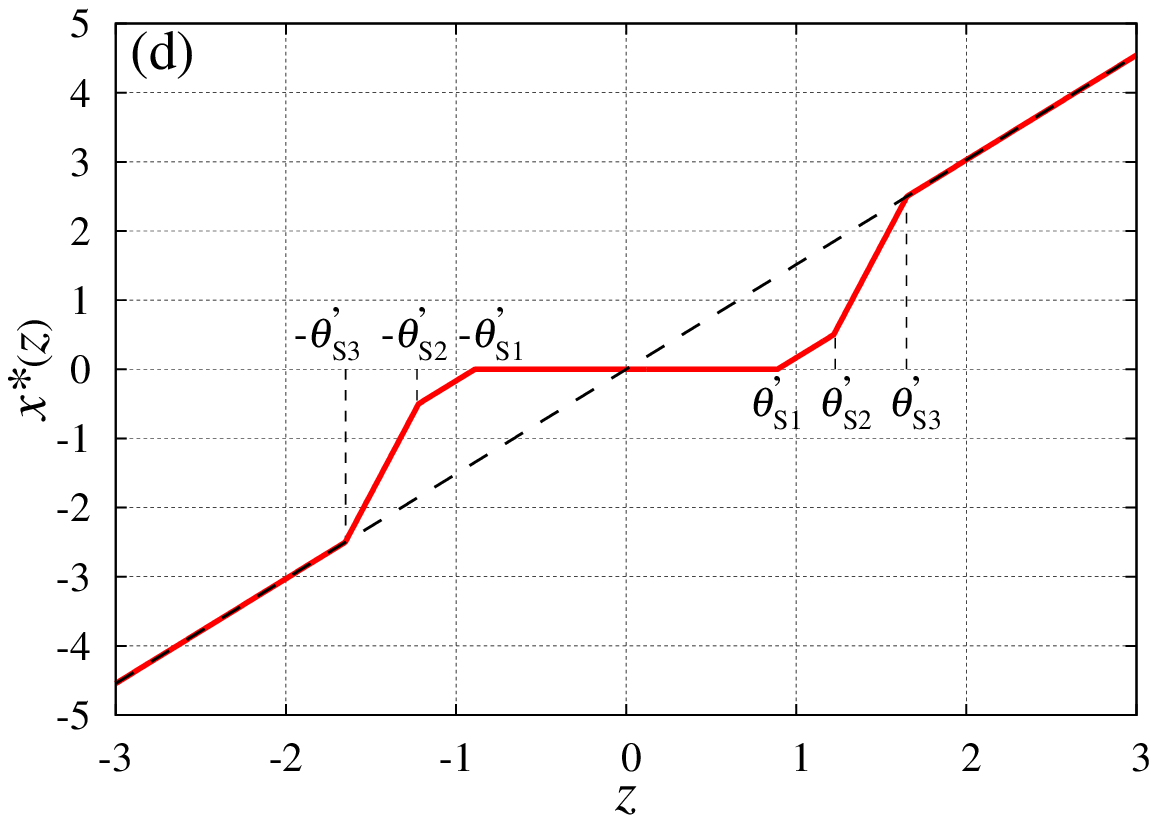}
\end{center}
\caption{Behaviour of the maximizers of the single-body problem at
$\alpha=1$ for
(a) $\ell_1$ ($\eta=1$), (b) elastic net ($\eta_1=1,~\eta_2=0.5$),
(c) $\ell_0$ ($\eta=1$), and (d) SCAD regularization ($\eta=1,~a=5,~\lambda=0.5$).
The dashed diagonal lines of gradient 1 are the maximizers under no regularization, and
the threshold $\theta^\prime$ denotes $\sqrt{2}\theta$.}
\label{fig:single-body}
\end{figure}

\subsection{$\ell_1$ regularization}

In the $\ell_1$ regularization
$r(\bm{x};\eta)=\eta||\bm{x}||_1=\eta\sum_i|x_i|$,
the maximizer of the single-body problem \eref{eq:f_pi_def}
is given by
\begin{eqnarray}
x^*_{\ell_1}(z;\hat{Q},\hat{\chi})=\left\{
\begin{array}{ll}
(h^{\rm RS}(z;\hat{\chi})-\eta{\rm sgn}(z))\slash\hat{Q} &
 (|h^{\rm RS}(z;\hat{\chi})|>\eta) \\
0 & ({\rm otherwise})
\end{array}\right.,
\label{eq:x_onebody}
\end{eqnarray}
where ${\rm sgn}(z)$ denotes the sign of $z$
and is $0$ when $z=0$.
Figure \ref{fig:single-body} (a) shows the behaviour of $x^*_{\ell_1}$
at $\alpha=1$ and $\eta=1$.
Setting $\theta_1=\eta\slash\sqrt{2\hat{\chi}}$,
the fraction of non-zero components 
is given by the probability that the solution of the RS single-body problem
\eref{eq:x_onebody} is non-zero:
$\hat{\rho}={\rm erfc}(\theta_1)$,
where
\begin{eqnarray}
{\rm erfc}(a)=\frac{2}{\sqrt{\pi}}\int_{a}^{\infty} dze^{-z^2},
\end{eqnarray}
and
\begin{eqnarray}
\pi_{\ell_1}=\frac{\hat{\chi}}{\hat{Q}}\Big\{(1+2\theta_1^2)\hat{\rho}-\frac{2\theta_1}{\sqrt{\pi}}e^{-\theta_1^2}\Big\}.
\end{eqnarray}
The regularization-dependent saddle point equations are given by
\begin{eqnarray}
\chi&=\frac{\hat{\rho}}{\alpha\hat{Q}}\label{eq:chi_L1}\\
Q&=\frac{\hat{\chi}\hat{\tau}_1}{\alpha\hat{Q}^2},
\label{eq:Q_L1}
\end{eqnarray}
where
\begin{eqnarray}
\hat{\tau}_1=(1+2\theta_1^2)\hat{\rho}-\frac{2\theta_1}{\sqrt{\pi}}e^{-\theta_1^2}.
\end{eqnarray}
From \eref{eq:RS_chih} and \eref{eq:RS_Qh},
the solutions of the saddle point equations \eref{eq:chi_L1}
and \eref{eq:Q_L1}
can be derived as
\begin{eqnarray}
\chi&=\frac{\hat{\rho}}{\alpha-\hat{\rho}}\label{eq:l1_chi}\\
Q&=\frac{(m_y^2+\sigma_y^2)\hat{\tau}_1}{\alpha-\hat{\tau}_1}.
\end{eqnarray}
Substituting the saddle point equations, the free energy density and
the expectation of the regularization term are given by
\begin{eqnarray}
f&=\frac{\alpha\hat{\chi}}{2}+\alpha(\chi\hat{\chi}-Q\hat{Q})\\
\overline{r}&=\lim_{N\to\infty}\frac{\eta}{N}E_{\bm{y},\bm{A}}[||\hat{\bm{x}}(\bm{y},\bm{A})||_1]=\alpha(\chi\hat{\chi}-Q\hat{Q}),
\end{eqnarray}
respectively.
Hence, the training error is given by
\begin{eqnarray}
\overline{\rm err}_{\rm train}=\hat{\chi}.
\end{eqnarray}
AT instability appears when
\begin{eqnarray}
\frac{\hat{\rho}}{\alpha}>1,
\end{eqnarray}
which is outside the region of interest for physical parameters.
Equation \eref{eq:l1_chi} leads to the following expression for the GDF:
\begin{eqnarray}
{\rm df}=\frac{\hat{\rho}}{\alpha}=\delta.
\label{eq:df_L1}
\end{eqnarray}
This expression is consistent with the result in \cite{L1_GDF},
which verifies the validity of this RS analysis
for the derivation of GDF.


\subsection{Elastic net regularization}

Elastic net regularization, given by
\begin{eqnarray}
r(\bm{x};\eta_1,\eta_2)=\eta_1||\bm{x}||_1+\frac{\eta_2}{2}||\bm{x}||_2^2,
\end{eqnarray}
was developed to encourage the grouping effect,
which is not exhibited by $\ell_1$ regularization, and
to stabilize the $\ell_1$ regularization path \cite{elastic_net}.
Here, the coefficient $1\slash 2$ is introduced for mathematical convenience,
and $\eta_2=0$ and $\eta_1=0$ correspond to $\ell_1$ regularization and
$\ell_2$ regularization, respectively.

The solution of the effective single-body problem for elastic net regularization is given by
\begin{eqnarray}
x^*_{\rm en}(z;\hat{Q},\hat{\chi})=\left\{
\begin{array}{ll}
(h^{\rm RS}(z;\hat{\chi})-\eta_1{\rm sgn}(z))\slash(\hat{Q}+\eta_2) & (|h^{\rm RS}(z;\hat{\chi})|>\eta_1) \\
0 & ({\rm otherwise})
\end{array}\right. .
\end{eqnarray}
The behaviour of this solution is shown in Fig. \ref{fig:single-body} (b) for
$\alpha=1,~\eta_1=1$, $\eta_2=0.5$,
and
\begin{eqnarray}
 \pi_{\rm en}=\frac{\hat{\chi}}{\hat{Q}+\eta_2}\Big\{(1+2\theta_{\rm en}^2)
{\rm erfc}(\theta_{\rm en})-\frac{2\theta_{\rm
en}}{\sqrt{\pi}}e^{-\theta_{\rm en}^2}\Big\},
\end{eqnarray}
where $\theta_{\rm en}=\eta_1\slash\sqrt{2\hat{\chi}}$.
The fraction of non-zero components 
is given by $\hat{\rho}={\rm erfc}(\theta_{\rm en})$, and
the regularization-dependent saddle point equations are given by
\begin{eqnarray}
Q&=\frac{\hat{\chi}}{\alpha(\hat{Q}+\eta_2)^2}\Big\{(1+2\theta_{\rm
 en}^2)\hat{\rho}-\frac{2\theta_{\rm
 en}}{\sqrt{\pi}}e^{-\theta_{\rm en}^2}\Big\}\\
\chi&=\frac{\hat{\rho}}{\alpha(\hat{Q}+\eta_2)}.
\label{eq:chi_saddle_en}
\end{eqnarray}
At the saddle point, the free energy density and the expectation of the
regularization term can be simplified as
\begin{eqnarray}
f&=\frac{\alpha\hat{\chi}}{2}+\alpha(\chi\hat{\chi}-Q\hat{Q})
-\frac{\alpha\eta_2Q}{2}\\
\nonumber
\overline{r}&=\lim_{N\to\infty}\frac{1}{N}E_{\bm{y},\bm{A}}\Big[\eta_1||\bm{\hat{x}}||_1+\frac{\eta_2}{2}||\bm{\hat{x}}||_2\Big]\\
&=\alpha\{\hat{\chi}\chi-(\hat{Q}+\eta_2)Q\}+\frac{\alpha\eta_2 Q}{2},
\end{eqnarray}
respectively.
Hence, the training error is given by
\begin{eqnarray}
\overline{\rm err}_{\rm train}=\hat{\chi}.
\end{eqnarray}
AT instability arises when
\begin{eqnarray}
\frac{\hat{\rho}}{\alpha}>\Big(\frac{\hat{Q}+\eta_2}{\hat{Q}}\Big)^2;
\end{eqnarray}
the right-hand side is
always greater than $1$ because $\eta_2\geq  0$ and $\hat{Q}>0$.
Therefore, the RS solution is always stable
under symmetry breaking perturbations
in the physical parameter region $\alpha>\hat{\rho}$.

From \eref{eq:chi_saddle_en},
the GDF for elastic net regularization is given by
\begin{eqnarray}
{\rm df}=\frac{\hat{\rho}}{\alpha}-\chi\eta_2=\frac{\delta\hat{Q}}{\hat{Q}+\eta_2},
\label{eq:df_en}
\end{eqnarray}
which reduces to the GDF for $\ell_1$ regularization at $\eta_2=0$.
An unbiased estimator of
the GDF for one instance of $\bm{A}$
is derived in \cite{Zou_DThesis}
as
\begin{eqnarray}
\hat{\rm df}(\bm{A})=\frac{1}{M}{\rm Tr}(\bm{A}_{{\cal A}}(\bm{A}_{{\cal A}}^{\rm T}\bm{A}_{{\cal A}}+\eta_2\bm{I}_{|{\cal A}|})^{-1}\bm{A}_{{\cal A}}^{\rm T}),
\end{eqnarray}
where ${\cal A}$ is the set of indices of non-zero components,
and the columns $\{\bm{A}_i|i\in{\cal A}\}$
constitute the submatrix $\bm{A}_{\cal A}$.
The number of the components of ${\cal A}$ is denoted by
$|{\cal A}|$.
Our expression \eref{eq:df_en} for df
corresponds to the typical value (or the expectation)
of $\hat{\rm df}(\bm{A})$
for a Gaussian random matrix $\bm{A}$.
The physical implications suggested by the cavity method \cite{beyond,MezardMontanari},
which is complementary to
the replica method, supports the correspondence relationship between $\hat{Q}$ in the replica method
and 
the Gram matrix of $\bm{A}$.
This correspondence indicates that our RS analysis is valid for the
derivation of GDF under elastic net regularization.

As shown in \eref{eq:df_en},
the GDF for elastic net regularization
deviates from $\delta=\hat{\rho}\slash\alpha$.
The $\ell_2$ regularization term in elastic net regularization
changes the variance of the non-zero components from $\hat{Q}^{-1}$ to
$(\hat{Q}+\eta_2)^{-1}$. Hence,
the effective fraction of the non-zero components measured by $\chi\slash \hat{Q}^{-1}$
does not coincide with $\delta$.
By defining the rescaled estimates of the single-body problem as
$x^{*{\rm res}}_{\rm en}=(1+\eta_2\slash\hat{Q})x^*_{\rm en}$,
the corresponding variance is reduced to $\chi^{\rm res}=\hat{\rho}\slash(\alpha\hat{Q})$
from \eref{eq:chi_single_body}, and
this gives ${\rm df}=\delta$.
This rescaling corresponds to that shown in \cite{elastic_net}, 
which was introduced to cancel out the shrinkage caused by $\ell_2$ regularization and improve the prediction performance.

Taking the limit as $\eta_1\to 0$,
the GDF for $\ell_2$ regularization
can be obtained where the estimate is not sparse.
The solution of the effective single-body problem is
given by
\begin{eqnarray}
x^*_{\ell_2}(z;\hat{Q},\hat{\chi})=\frac{h^{\rm RS}(z;\hat{\chi})}{\hat{Q}+\eta_2},
\end{eqnarray}
and the function $\pi$ is given by
\begin{eqnarray}
\pi_{\ell_2}=\frac{\hat{\chi}}{\hat{Q}+\eta_2}.
\end{eqnarray}
This expression leads to the following GDF:
\begin{eqnarray}
{\rm df}=\frac{\hat{Q}}{\hat{Q}+\eta_2},
\label{eq:df_l2}
\end{eqnarray}
which corresponds to the limit as $\delta\to 1$ 
of the elastic net regularization.
An unbiased estimator of
the GDF for one instance of $\bm{A}$ is proposed as \cite{Hastie_Tibshirani1990}
\begin{eqnarray}
\hat{\rm df}(\bm{A})=\frac{1}{M}{\rm Tr}\bm{A}(\bm{A}^{\rm T}\bm{A}+\eta_2\bm{I}_N)^{-1}\bm{A}^{\rm T}.
\end{eqnarray}
\Eref{eq:df_l2} corresponds to 
the expectation of $\hat{\rm df}(\bm{A})$
for a Gaussian random matrix $\bm{A}$.


\subsection{$\ell_0$ regularization}

The $\ell_0$ regularization is expressed by
$r(\bm{x};\eta)=\eta||\bm{x}||_0=\eta\sum_i|x_i|_0$,
which corresponds to the number of non-zero components in $\bm{x}$.
The solution to the single-body problem
for $\ell_0$ regularization is given by
\begin{eqnarray}
x_{\ell_0}^*(z;\hat{Q},\hat{\chi})=\left\{
\begin{array}{ll}
h^{\rm RS}(z;\hat{\chi})\slash\hat{Q} & (|h^{\rm RS}(z;\hat{\chi})|>\sqrt{2\hat{\chi}}\theta_0)\\
0 & ({\rm otherwise})
\end{array} \right.,
\label{eq:RS_1body}
\end{eqnarray}
where $\theta_0=\sqrt{\eta\hat{Q}\slash\hat{\chi}}$,
and by setting the fraction of non-zero components
to $\hat{\rho}={\rm erfc}(\theta_0)$, we can derive
\begin{eqnarray}
\pi_{\ell_0
}=\frac{\hat{\chi}}{\hat{Q}}\Big\{\frac{2\theta_0}{\sqrt{\pi}}
e^{-\theta_0^2}+(1-2\theta_0^2)\hat{\rho}\Big\}.
\end{eqnarray}
Figure \ref{fig:single-body} (c)
shows the $z$-dependence of the maximizer $x^*_{\ell_0}$
at $\alpha=1$ and $\eta=1$.
The regularization-dependent saddle point equations \eref{eq:RS_chi_gen}--\eref{eq:RS_Q_gen} are
given by
\begin{eqnarray}
\chi&=\frac{1}{\alpha\hat{Q}}\Big\{\frac{2\theta_0}{\sqrt{\pi}}
e^{-\theta_0^2}+\hat{\rho}\Big\}\label{eq:l0_chi}\\
Q&=\frac{\hat{\chi}}{\alpha\hat{Q}^2}\Big\{\frac{2\theta_0}{\sqrt{\pi}}
e^{-\theta_0^2}+\hat{\rho}\Big\},\label{eq:l0_Q}
\end{eqnarray}
and have two solutions: finite $\chi$ and $Q$
and infinite $\chi$ and $Q$.
We denote the finite and infinite solutions
as ${\rm S}_1=\{\chi_1,Q_1\}$ and ${\rm S}_2=\{\chi_2=\infty,Q_2=\infty\}$,
respectively.
Using \eref{eq:RS_chih} and \eref{eq:RS_Qh},
the finite solution can be simplified
as
\begin{eqnarray}
\chi_1&=\frac{\hat{\rho}+\omega}
{\alpha-(\hat{\rho}+\omega)}\label{eq:l0_chi_finite}\\
Q_1&=(m_y^2+\sigma_y^2)\chi_1,\label{eq:l0_Q_finite}
\end{eqnarray}
where
\begin{eqnarray}
\omega=\int Dz|z|\delta(|z|-\sqrt{2}\theta_0)=\frac{2\theta_0}{\sqrt{\pi}}e^{-\theta_0^2}.
\label{eq:w_hat}
\end{eqnarray}
By definition, $\chi_1$ and $Q_1$ should be positive,
and so \eref{eq:l0_chi_finite}--\eref{eq:l0_Q_finite}
are only valid when $\alpha>\hat{\rho}+\omega$.
According to a local stability analysis
of \eref{eq:l0_chi} around $1\slash\chi=0$, solution ${\rm S}_2$ is 
a locally stable
solution of the RS saddle point equation
when
$\alpha<\hat{\rho}+\omega$,
where as it is unstable when $\alpha>\hat{\rho}+\omega$.
Therefore, the stable
solution of the RS saddle point equation
changes from ${\rm S}_1$
to ${\rm S}_2$
at $\alpha=\hat{\rho}+\omega$.
Note that the
stability discussed here refers to the RS solution, and
does not relate to 
AT instability.

The free energy density is simplified
by substituting the saddle point equations as
\begin{eqnarray}
f=\frac{\alpha\hat{\chi}}{2}+\eta\hat{\rho}.
\label{eq:f_l0}
\end{eqnarray}
The second term of \eref{eq:f_l0} corresponds to the
expectation of the regularization term,
and so the training error can be derived as
\begin{eqnarray}
\overline{\rm err}_{\rm train}=\hat{\chi}.
\end{eqnarray}
The GDF is given by
\begin{eqnarray}
{\rm df}=
\left\{
\begin{array}{ll}
\delta+\displaystyle\frac{\omega}{\alpha} & {\rm for~solution}~
{\rm S}_1\\
1 & {\rm for~solution}~
{\rm S}_2
\end{array}
\right..
\label{eq:l0_GDF}
\end{eqnarray}
The term $\omega$, given by \eref{eq:w_hat}, in the GDF
originates from the discontinuity of the single-body problem
at the threshold $\sqrt{2}\theta_0$, as shown in
\fref{fig:single-body} (c).
In addition to the fluctuation generated by the non-zero components,
this discontinuity induces fluctuations in the system
and increases the GDF from $\delta$.

Under $\ell_0$ regularization,
AT instability always appears,
but the estimated GDF under the RS assumption can be regarded as an approximation of the true value of the GDF, as shown in Sec.~\ref{sec:l0_exhaustive}.
Our calculations based on the one-step RSB assumption
indicate that the form of the GDF, as
the fraction of non-zero components plus the 
discontinuity term, is unchanged, although the
values of these two terms does change (unreported).


\subsection{SCAD regularization}

SCAD regularization is a non-convex sparse regularization
in which the estimator has the desirable properties of being unbiased, sparse, and continuous \cite{SCAD}.
Mathematically, the SCAD estimator is asymptotically equivalent
to the oracle estimator \cite{SCAD,SCAD2}.
SCAD regularization is given by
\begin{eqnarray}
r(x;\eta)=\left\{\begin{array}{ll}
\eta\lambda|x| & (|x|\leq \lambda) \\
-\eta\Big\{\displaystyle\frac{x^2-2a\lambda|x|+\lambda^2}{2(a-1)}\Big\} & (\lambda<|x|\leq a\lambda) \\
\displaystyle\frac{\eta(a+1)\lambda^2}{2} & |x|>a\lambda
\end{array}
\right.,
\end{eqnarray}
where $\lambda$ and $a$ are parameters that
control the form of the regularization.
The maximizer of the single-body problem for SCAD regularization is
given by
\begin{eqnarray}
\nonumber
x^*_{\rm S}(z;\hat{Q},\hat{\chi})=\left\{\begin{array}{ll}
\displaystyle\frac{h^{\rm RS}(z;\hat{\chi})-\lambda\eta{\rm sgn}(z)}{\hat{Q}} &
 (\lambda\eta<|h^{\rm RS}(z;\hat{\chi})|\leq \lambda(\hat{Q}+\eta))\\
\displaystyle\frac{h^{\rm RS}(z;\hat{\chi})(a-1)-{a\lambda\eta{\rm sgn}(z)}}{\hat{Q}(a-1)-\eta} &
 (\lambda(\hat{Q}+\eta)<|h^{\rm RS}(z;\hat{\chi})|\leq a\lambda\hat{Q})\\
\displaystyle\frac{{h}^{\rm RS}(z;\hat{\chi})}{\hat{Q}} & (|h^{\rm RS}(z;\hat{\chi})|>a\lambda\hat{Q})\\
0& ({\rm otherwise})
\end{array}
\right..\\
\label{eq:single_body_SCAD}
\end{eqnarray}
Figure \ref{fig:single-body} (d) shows an example of the
behaviour of the maximizer $x^*_{\rm S}$
at $a=5,\lambda=0.1$, and $\eta=1$,
where three thresholds are given by
$\theta_{S1}=\lambda\eta\slash\sqrt{2\hat{\chi}}$,
$\theta_{S2}=\lambda(\hat{Q}+\eta)\slash\sqrt{2\hat{\chi}}$, and
$\theta_{S3}=a\lambda\hat{Q}\slash\sqrt{2\hat{\chi}}$.
The threshold $\theta_{{\rm S}1}$ gives the
fraction of non-zero components as
$\hat{\rho}={\rm erfc}(\theta_{S1})$.
Between the thresholds $\theta_{{\rm S}1}$ and $\theta_{{\rm S}2}$,
and beyond the third threshold $\theta_{{\rm S}3}$,
the estimate $x^*_{\rm S}$
behaves like the $\ell_1$ and $\ell_0$
estimates, respectively.
Between $\theta_{{\rm S}2}$ and $\theta_{{\rm S}3}$,
the estimate transits linearly between the $\ell_1$ and $\ell_0$ estimates.

The function $\pi$ for SCAD regularization is derived as 
\begin{eqnarray}
\pi_{\rm
 S}=\pi_1+\pi_2+\pi_3+\frac{\eta\lambda^2\pi_4}{a-1}-\eta(a+1)\lambda
 ^2{\rm erfc(\theta_{{\rm S}3})},
\end{eqnarray}
where
\begin{eqnarray}
\pi_1&\!=\!\frac{\hat{\chi}}{\hat{Q}}\Big[\!-\!\frac{2\theta_{S1}}{\sqrt{\pi}}\Big(\!e^{-\theta_{S1}^2}\!+\!(\frac{\hat{Q}\!-\!\eta}{\eta})e^{-\theta_{S2}^2}\!\Big)
\!+\!(1\!+\!2\theta_{S1}^2)\{\hat{\rho}\!-\!{\rm erfc}(\theta_{S2}\!)\}\!\Big]\\
\nonumber
\pi_2&=\frac{\hat{\chi}}{\hat{Q}-\frac{\eta}{a-1}}\Big[\frac{2}{\sqrt{\pi}}\Big\{\Big(\theta_{S2}-\frac{2\theta_{S3}\eta}{\hat{Q}(a-1)}\Big)e^{-\theta_{S2}^2}
-\Big(1-\frac{2\eta}{\hat{Q}(a-1)}\Big)\theta_{S3}e^{-\theta_{S3}^2}\Big\}\\
&+\Big\{1+2\Big(\frac{\eta\theta_{S3}}{\hat{Q}(a-1)}\Big)^2\Big\}\pi_4\Big]\\
\pi_3&=\frac{\hat{\chi}}{\hat{Q}}\Big[\frac{2\theta_{S3}}{\sqrt{\pi}}e^{-\theta_{S3}^2}+{\rm erfc}(\theta_{S3})\Big]\\
\pi_4&={\rm erfc}(\theta_{S2})-{\rm erfc}(\theta_{S3}).
\end{eqnarray}
The regularization-dependent saddle point equations are given by
\begin{eqnarray}
Q&=\frac{1}{\alpha}\left\{\frac{\pi_1}{\hat{Q}}+\frac{\pi_2}{\hat{Q}-\frac{\eta}{a-1}}+\frac{\pi_3}{\hat{Q}}\right\}\\
\chi&=\frac{1}{\alpha\hat{Q}}\Big[\hat{\rho}+\frac{\frac{\eta}{a-1}}{\hat{Q}-\frac{\eta}{a-1}}\pi_4\Big],
\end{eqnarray}
and the expectation of the regularization term is given by
\begin{eqnarray}
\nonumber
\overline{r}&=\alpha\chi\hat{\chi}-\pi_1-\Big\{1+\frac{\frac{\eta}{a-1}}{2(\hat{Q}-\frac{\eta}{a-1})}\Big\}\pi_2
-\pi_3-\frac{\eta\lambda^2}{2(a-1)}\pi_4+\frac{\eta(a+1)\lambda^2}{2}
{\rm erfc}(\theta_{{\rm S}3}).\\
\end{eqnarray}
Substituting these equations into the free energy density, we get
\begin{eqnarray}
\overline{\rm err}_{\rm train}=\hat{\chi}.
\end{eqnarray}
The AT instability condition is given by
\begin{eqnarray}
\frac{1}{\alpha(1+\chi)^2}\Big[\frac{\hat{\rho}}{\hat{Q}^2}+\left\{\left(\hat{Q}\!-\!\frac{\eta}{a-1}\right)^
{-2}\!\!\!\!\!-\frac{1}{\hat{Q}^{2}}\right\}\pi_4\Big]>1,
\end{eqnarray}
which reduces to that for $\ell_1$ regularization as $a\to\infty$.

There are three solutions of $\{Q,\chi\}$:
${\rm S}_1=\{Q=Q_1<\infty,\chi=\chi_1<\infty\}$,
${\rm S}_2=\{Q=Q_2<\infty,\chi=\infty\}$,
and ${\rm S}_3=\{Q=\infty,\chi=\infty\}$.
For sufficiently large $a$,
the finite solution ${\rm S}_1$
is a locally stable solution of the RS saddle point equation
when
\begin{eqnarray}
\alpha>\hat{\rho}+\frac{\eta\slash\{\hat{Q}(a-1)\}}{1-\eta\slash\{\hat{Q}(a-1)\}}\pi_4.
\label{eq:SCAD_S1_stability}
\end{eqnarray}
Beyond the range of \eref{eq:SCAD_S1_stability},
the stable RS solution is replaced by ${\rm S}_2$.
For sufficiently small $\eta$,
the stable RS solution
can switch from ${\rm S}_2$ to ${\rm S}_3$
depending on the SCAD parameter,
but this is not important in estimating the GDF,
because both solutions
give the same GDF value.
The GDF for SCAD regularization can be summarized as
\begin{eqnarray}
{\rm df}=\left\{\begin{array}{ll}
\delta+\displaystyle\frac{\frac{\eta}{a-1}}{\alpha(\hat{Q}-\frac
 {\eta}{a-1})}\pi_4 & {\rm for~solution~S}_1\\
1 & {\rm for~solutions~S}_2{\rm~and~S}_3
\end{array}
\right..
\label{eq:SCAD_df}
\end{eqnarray}
As $a\to\infty$, we get $\pi_4\to 0$
and solution ${\rm S}_1$ is always
a stable RS solution,
satisfying \eref{eq:SCAD_S1_stability}; hence,
the GDF reduces to that for $\ell_1$ regularization.
The second term of the GDF for solution ${\rm S}_1$
arises from the weight between the thresholds
$\theta_{{\rm S}2}$ and $\theta_{{\rm S}3}$.
The manner of assigning the non-zero components to this
transient region between the $\ell_1$ and $\ell_0$ estimates
increases the fluctuation in the system,
and the GDF does not coincide with $\delta$.

We note the pathology of 
solution ${\rm S}_3$ under the RS assumption.
As shown in the solution to the single-body problem \eref{eq:single_body_SCAD}
(Fig. \ref{fig:single-body} (d)),
the magnitude relation $\theta_{{\rm S}1}\leq\theta_{{\rm
S}2}\leq\theta_{{\rm S}3}$ should hold.
However, the $Q,\chi\to\infty$ solution leads to $\theta_{{\rm S}3}\to0$
with finite $\theta_{{\rm S}1}=\theta_{{\rm S}2}$.
Solution ${\rm S}_3$ appears
in the region where
AT instability appears,
and so this non-physical phenomenon is considered
to be caused by an inappropriate RS assumption.
Hence, we must construct the RSB solution to correctly
describe the GDF corresponding to solution ${\rm S}_3$.

\section{Parameter dependence of GDF and prediction error}
\label{sec:GDF}

\begin{figure}
\begin{center}
\includegraphics[width=3.5in]{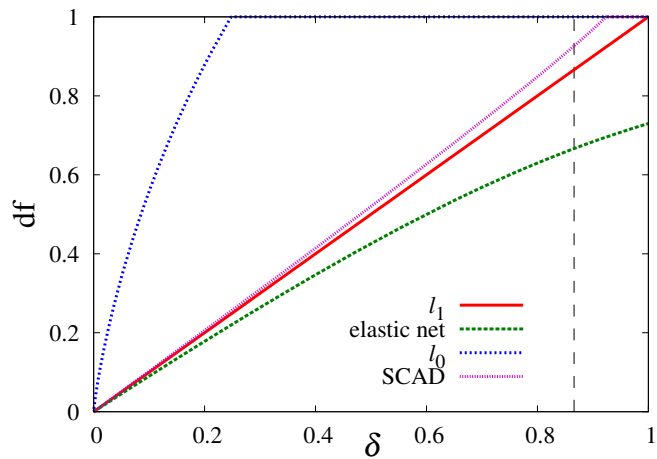}
\end{center}
\caption{$\delta$-dependence of GDF
for $\ell_1$, elastic net, $\ell_0$, and SCAD regularization
at $\alpha=0.5$, $m_y=0$, and $\sigma_y^2=1$.
The parameters for elastic net and SCAD regularization are $\eta_2=0.1$,
$a=8$, and $\lambda=1$,
and the vertical dashed line indicates the
appearance of the AT instability for SCAD regularization,
$\delta=0.866$.
The $\ell_1$ result corresponds to the line ${\rm df}=\delta$.}
\label{fig:GDF}
\end{figure}

Figure \ref{fig:GDF} illustrates the $\delta$-dependence of the GDF
for $\ell_1$, the elastic net with $\eta_2=0.1$, $\ell_0$,
and SCAD regularization with $a=8$ and $\lambda=1$
at $\alpha=0.5$, $m_y=0$, and $\sigma_y^2=1$.
At each point of $\delta$,
the regularization parameters
$\eta$ for $\ell_1$, $\ell_0$ and SCAD regularization
and $\eta_2$ for elastic net regularization
are controlled such that $\delta=\hat{\rho}\slash\alpha$.
Under $\ell_1$ regularization, the GDF is always equal to $\delta$,
as shown in \eref{eq:df_L1}.
In elastic net regularization,
the GDF is less than $\delta$
as the $\ell_2$ parameter $\eta_2$ increases.
For $\ell_0$ regularization,
the RS solution ${\rm S}_1$ 
is unstable at $\delta>0.248$ in this parameter region,
and is replaced by solution ${\rm S}_2$, which
gives ${\rm df}=1$.
In SCAD regularization,
the solution ${\rm S}_1$ loses local stability
within the RS assumption at
$\delta>0.924$, and
AT instability 
appears before the RS solution ${\rm S}_1$ becomes unstable
at $\delta>0.866$
(denoted by the dashed vertical line in \fref{fig:GDF}.)

\begin{figure}
\begin{center}
\includegraphics[width=4in]{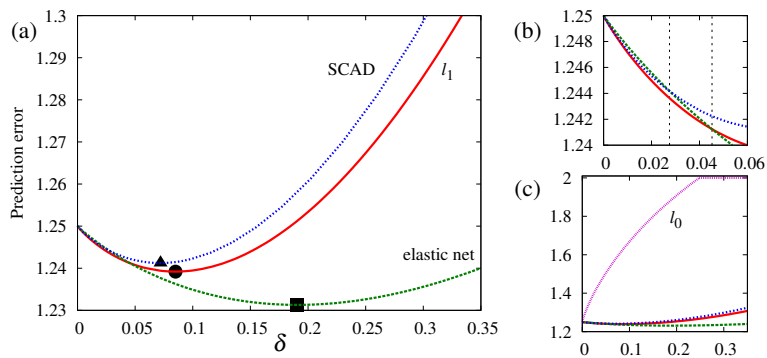}
\end{center}
\caption{(a) $\delta$-dependence of the prediction error
$\overline{\rm err}_{\rm pre}$
for $\ell_1$, elastic net, and SCAD regularization
at $\alpha=0.5$, $m_y=0.5$, and $\sigma_y^2=1$.
The parameters for elastic net and SCAD regularization are $\eta_2=0.1$, $a=8$, and $\lambda=1$.
(b) Region where the magnitude relationship
between each regularization changes.
(c) Prediction error for $\ell_0$ regularization.
}
\label{fig:pre_error}
\end{figure}

Figure \ref{fig:pre_error} shows the prediction error \eref{eq:pre_error}
for the same parameter region as \fref{fig:GDF}.
At $\sigma_y^2=1$,
the prediction error is equivalent to the expectation of AIC.
In the entire range of $\delta$ shown in \fref{fig:pre_error} (a),
the RS solutions for
$\ell_1$, elastic net, and SCAD regularization are stable
under symmetry breaking perturbations.
Thus, we can identify the value of $\delta$
that minimizes the prediction error 
for each regularization.
In this case,
the models with $\delta=0.085$ (denoted by \textbullet), $\delta=0.170$
($\blacksquare$), and $\delta=0.072$ ($\blacktriangle$)
are selected for $\ell_1$, elastic net, and SCAD regularization,
respectively.
In the current problem setting,
sparse estimation with SCAD regularization 
minimizes the prediction error 
within the RS region when the mean of the data is sufficiently small.
To identify the appropriate model
using RS analysis, it is useful to standardize the data.
As shown in \fref{fig:pre_error} (b),
the magnitude of the prediction errors
at $\delta<\delta_1=0.028$, $\delta_1<\delta<\delta_2=0.045$,
and $\delta>\delta_2$ runs 
in descending order as elastic net$>$SCAD$>\ell_1$,
SCAD$>$elastic net$>\ell_1$,
and SCAD$>\ell_1>$elastic net, respectively.
The estimates $\hat{\bm{x}}$
have different supports depending on the regularization,
even when the regularization parameters are controlled to
give a certain value of $\delta$.
A comparison of the prediction errors
within the framework of RS analysis
guides the choice of regularization
for each value of $\delta$.

The prediction error for $\ell_0$ regularization
under the RS assumption is shown in
\fref{fig:pre_error} (c) alongside those for other regularization types.
The RS prediction error is minimized 
at $\delta=0$.
This indicates that the appropriate model
under RS analysis has
a non-zero component of $O(1)$.
Our analysis assumes that
the number of non-zero components is $O(N)$;
hence, the derived model selection criterion
cannot identify the appropriate model
in the current problem setting
for $\ell_0$ regularization.

\section{Numerical calculation of GDF using belief propagation algorithm}
\label{sec:BP}

\subsection{Belief propagation algorithm for sparse regularization}
The correspondence between replica analysis
and the belief propagation (BP) algorithm
suggests that the typical properties of BP fixed points at the large-system-size limit
can be described by
the RS saddle point \cite{beyond,MezardMontanari}.
Thus, we may expect that the numerically obtained
GDF will be consistent with the RS analysis at finite system sizes
using the BP algorithm.
For the ordinary least squares with a regularization that
can be written as \eref{eq:lasso},
a tentative estimate of the $i$-th component at step $t$,
denoted by $\hat{x}_i^{(t)}$, is given by
the solution to the single-body problem \eref{eq:one_body}
with the substitutions $\hat{Q}\to\hat{Q}_i^{(t)}$ and $h^{\rm
RS}(z;\hat{\chi})\to h_i^{(t)}$ \cite{AMP,DMM2,GAMP}, where
\begin{eqnarray}
h_i^{(t)}&=\hat{x}_i^{(t-1)}\sum_{\mu=1}^M\frac{A_{\mu
 i}^2}{1+{\sigma_\mu^{(t-1)}}^2}+\sum_{\mu=1}^MA_{\mu
 i}R_\mu^{(t-1)}\\
\hat{Q}_i^{(t)}&=\sum_{\mu=1}^M\frac{A_{\mu i}^2}{1+{\sigma_\mu^{(t-1)}}^2},
\end{eqnarray}
and setting $\hat{\bm{y}}^{(t)}=\bm{A}\hat{\bm{x}}^{(t)}$,
\begin{eqnarray}
R_\mu^{(t)}&=\frac{y_\mu-\hat{y}_{\mu}^{(t)}}{1+{\sigma_\mu^{(t)}}^2}\\
{\sigma_\mu^{(t)}}^2&=\frac{1}{\alpha}\sum_iA_{\mu i}^2\chi_i^{(t)}.
\end{eqnarray}
The variable $\chi_i^{(t)}$ represents the variance of $x_i^{(t)}$, and its determination rule depends on the regularization.
For $\ell_1$ and elastic net regularization,
the variable is given by
\begin{eqnarray}
  \chi_i^{(t)}=\left\{\begin{array}{ll}
  \displaystyle\frac{1}{\hat{Q}_i^{(t)}} & {\rm for}~|h_i^{(t)}|>\eta\\
  0 & {\rm otherwise}
  \end{array}
  \right.
\end{eqnarray}
and 
\begin{eqnarray}
  \chi_i^{(t)}=\left\{\begin{array}{ll}
  \displaystyle\frac{1}{\hat{Q}_i^{(t)}+\eta_2} & {\rm for}~|h_i^{(t)}|>\eta_1\\
  0 & {\rm otherwise}
  \end{array}
  \right.,
\end{eqnarray}
respectively.
For these regularizations,
the GDF at the BP fixed point converges to
that given by RS analysis as the system size increases.
However, 
for these regularizations, the GDF can be calculated 
using least angle regression (LARS) \cite{LARS},
which has a lower computational cost than the BP algorithm.
Hence, there is no need to introduce the BP algorithm.
In the case of 
$\ell_0$ regularization,
the variable $\chi_i$ is given by
\begin{eqnarray}
  \chi_i^{(t)}=\left\{\begin{array}{ll}
  \displaystyle\frac{1}{\hat{Q}_i^{(t)}} & {\rm for}~|h_i^{(t)}|>\sqrt{2\eta\hat{Q}_i^{(t)}}\\
  0 & {\rm otherwise}
  \end{array}
  \right..
\end{eqnarray}
Unfortunately,
AT instability appears across the whole parameter region
for $\ell_0$ regularization,
and the BP algorithm does not converge.

For SCAD regularization,
no numerical method for 
the precise evaluation of GDF has been proposed.
As shown in the previous section,
SCAD regularization
gives a parameter region where the RS solution is stable.
Therefore, the BP algorithm is useful as a method of
numerically calculating the GDF for SCAD regularization.
The variable ${\chi}_i^{(t)}$
for SCAD regularization
is given by
\begin{eqnarray}
{\chi_i^{(t)}}&=\left\{\begin{array}{ll}
\displaystyle\frac{1}{\hat{Q}_i^{(t)}} & {\rm
 for~}\lambda\eta<|h^{(t)}_i|\leq\lambda(\hat{Q}_i^{(t)}+\eta)\mbox{ and }|h^{(t)}_i|>a\lambda\hat{Q}^{(t)}_i\\
\displaystyle\frac{1}{\hat{Q}_i^{(t)}-\frac{\eta}{a-1}} &
 \mbox{for }\lambda(\hat{Q}_i^{(t)}+\eta)<|h_i^{(t)}|\leq a\lambda\hat{Q}_i^{(t)}\\
0 & {\rm otherwise}
\end{array}
\right..
\end{eqnarray}
After updating the estimates $\hat{\bm{x}}^{(t)}$,
we can numerically evaluate the value of GDF using \eref{eq:def_df}
with the data estimates $\hat{\bm{y}}^{(t)}$.
To ensure convergence, appropriate damping is required at each update.

\begin{figure}
\begin{center}
\includegraphics[width=3in]{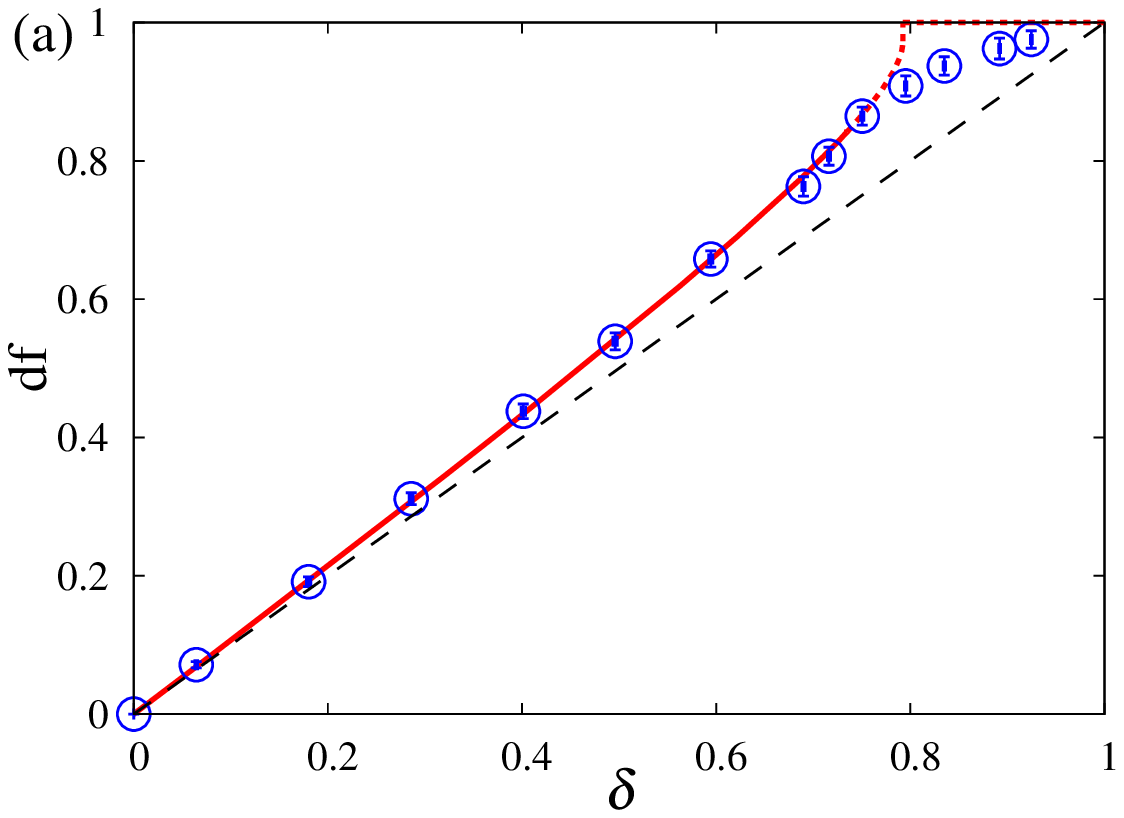}
\includegraphics[width=3in]{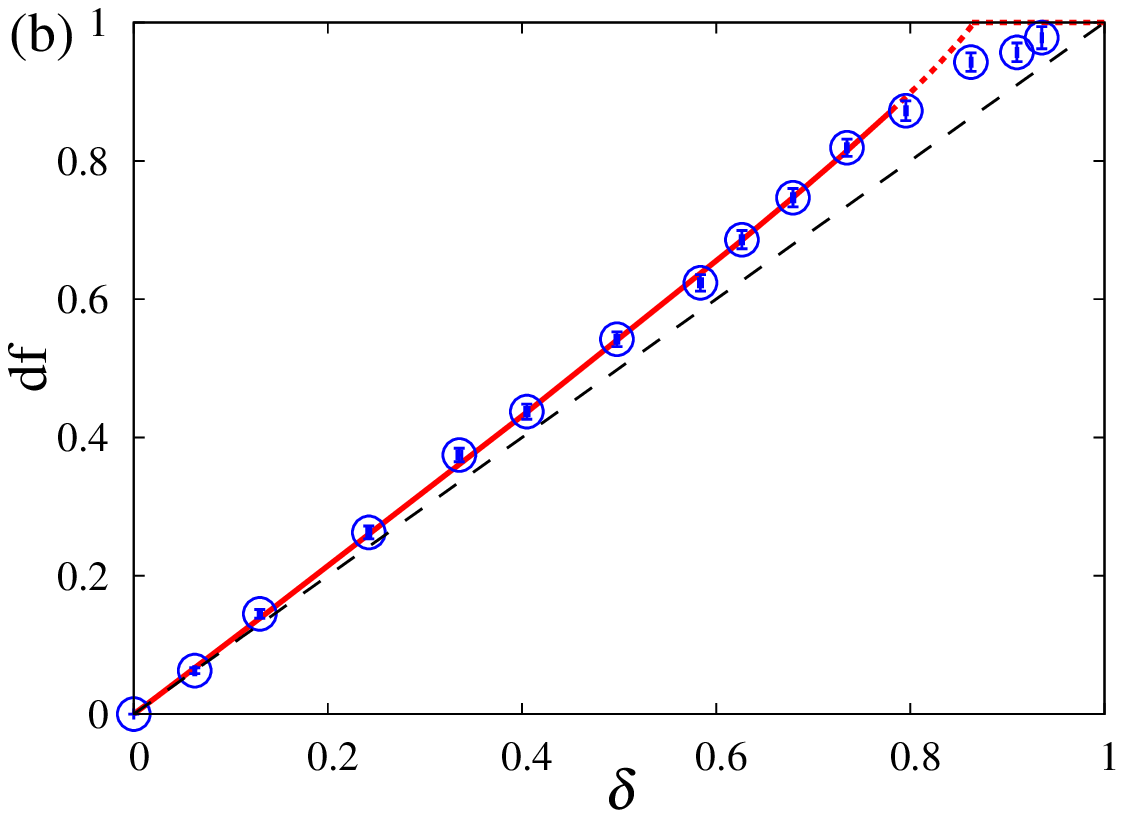}
\includegraphics[width=3in]{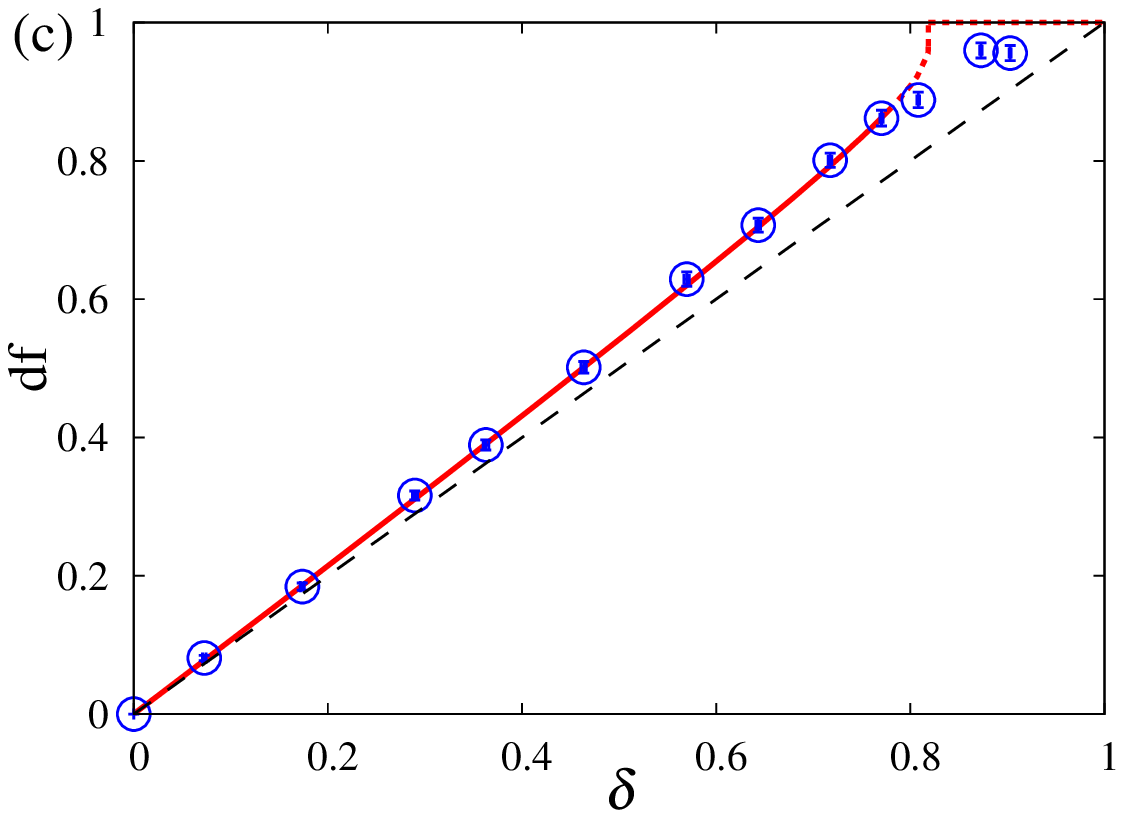}
\includegraphics[width=3in]{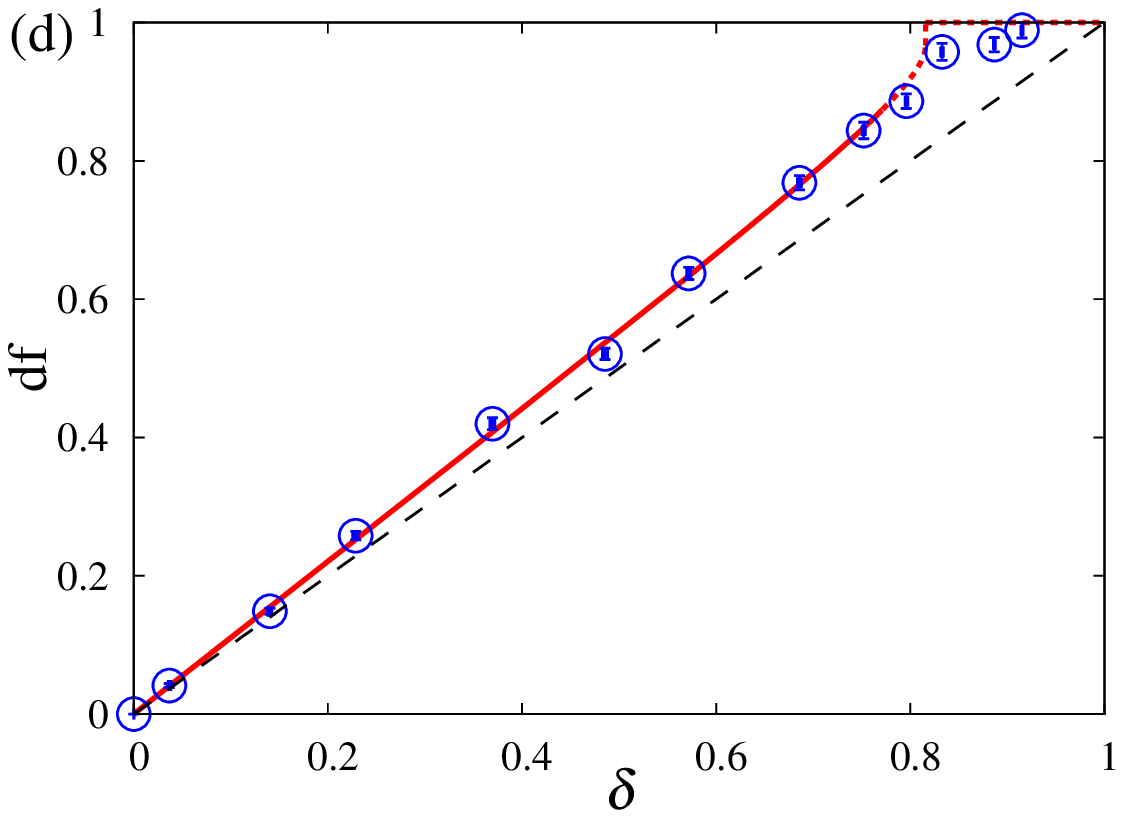}
\end{center}
\caption{Comparison between BP algorithm and RS analysis
for SCAD regularization at $N=200$, $m_y=0$, and $\sigma_y^2=1$ for
(a) $a=5$, $\lambda=1$ for $M=100$ ($\alpha=0.5$), (b) $a=8$, $\lambda=0.8$ for $M=100$ ($\alpha=0.5$), (c) $a=6$, $\lambda=0.9$ for $M=160$ ($\alpha=0.8$), and (d) $a=8$, $\lambda=0.7$ for $M=160$ ($\alpha=0.8$).
The BP results, averaged over 100 realizations of $\{\bm{y},\bm{A}\}$,
are denoted by circles.
The theoretical estimation of GDF by RS analysis is denoted by solid and dashed lines for the RS and RSB regions, respectively.
To provide a visual guide, the dashed line has a gradient of $1$.}
\label{fig:BP}
\end{figure}

As for the replica analysis, we apply the BP algorithm for
the case of Gaussian random data $\bm{y}$
and predictors $\bm{A}$.
Figure \ref{fig:BP} shows the
numerically calculated GDF given by the BP algorithm
at $N=200$, $m_y=0$, and $\sigma_y^2=1$.
The BP algorithm is updated until $|\hat{x}_i^{(t)}-\hat{x}_i^{(t-1)}|<10^{-10}$
for each component,
and the result is averaged over 100 realizations of $\{\bm{y},\bm{A}\}$.
The solid and dashed lines
represent the analytical results given by the replica method
for the RS and RSB regions, respectively.
In the RS regime, the numerically calculated GDF from the BP algorithm
coincides with that evaluated by the replica method.

\subsection{Perspective for other predictors}
The RS analysis
discussed so far has been applied to
Gaussian i.i.d. random predictors.
Its extension to other predictors
is not straightforward.
To check the generality of the 
GDF being given by the effective fraction of non-zero components
$(\chi\slash\hat{Q}^{-1})$
at the RS saddle point for other predictor matrices,
we resort to the BP algorithm.
The typical properties of $\chi_i^{(t)}$ and $\hat{Q}_i^{(t)}$ at
the BP fixed point denoted by
$\chi_i^*$ and $\hat{Q}_i^*$ are described in the replica method
by $\chi$ and $\hat{Q}$
of the RS saddle point 
at the large-system-size limit.
Therefore, it is reasonable to define
the effective fraction of non-zero components
at the BP fixed point as
\begin{eqnarray}
\delta^{\rm BP}_{\rm eff}=\frac{\overline{\frac{1}{N}\sum_i\chi_i^*(\bm{y},\bm{A})}}{\left(\overline{\frac{1}{N}\sum_i\hat{Q}^*_i(\bm{y},\bm{A})}\right)^{-1}},
\end{eqnarray}
where the overline represents the average over
$\bm{y}$ and $\bm{A}$.
If $\delta_{\rm eff}^{\rm BP}$ and the GDF from \eref{eq:def_df}
coincide at the BP fixed point,
it is considered that
the correspondence between GDF and the effective fraction of non-zero
components holds at the RS saddle point.
For $\ell_1$, elastic net, and SCAD regularization,
we examine the behaviour of GDF and $\delta_{\rm eff}^{\rm BP}$ under
two predictors \cite{elastic_net}
in a parameter region where the BP algorithm converges.
\begin{description}
\item[Example 1:] Gaussian predictors with pairwise correlation. The correlation between predictors
$\bm{A}_i$ and $\bm{A}_j$ is set to be $c^{|i-j|}$, and
the predictors are normalized such that $||\bm{A}_i||_2^2=1$.
\item[Example 2:] The predictors are generated as
\begin{eqnarray}
\bm{A}_i=\left\{\begin{array}{ll}
\bm{Z}_1+\bm{\epsilon}_i & {\rm for}~i=1,\cdots,T\\
\bm{Z}_2+\bm{\epsilon}_i & {\rm for}~i=K+1,\cdots,2T\\
\bm{Z}_3+\bm{\epsilon}_i & {\rm for}~i=2K+1,\cdots,3T\\
\bm{a}_i & {\rm for}~i=3T+1,\cdots,N\\
\end{array}
\right.,
\end{eqnarray}
where the components of the $M$-dimensional vectors
$\bm{Z}_1,\bm{Z}_2,\bm{Z}_3$,
$\{\bm{a}_i\}$,
and $\{\bm{\epsilon}_i\}$
are i.i.d. Gaussian random variables with mean zero and variance $1$,
and $T$ is a parameter that takes an integer value smaller than
$(N-1)\slash 3$.
The predictors are normalized such that $||\bm{A}_i||_2^2=1$.
\end{description}

\begin{figure}
\begin{center}
\includegraphics[width=2in]{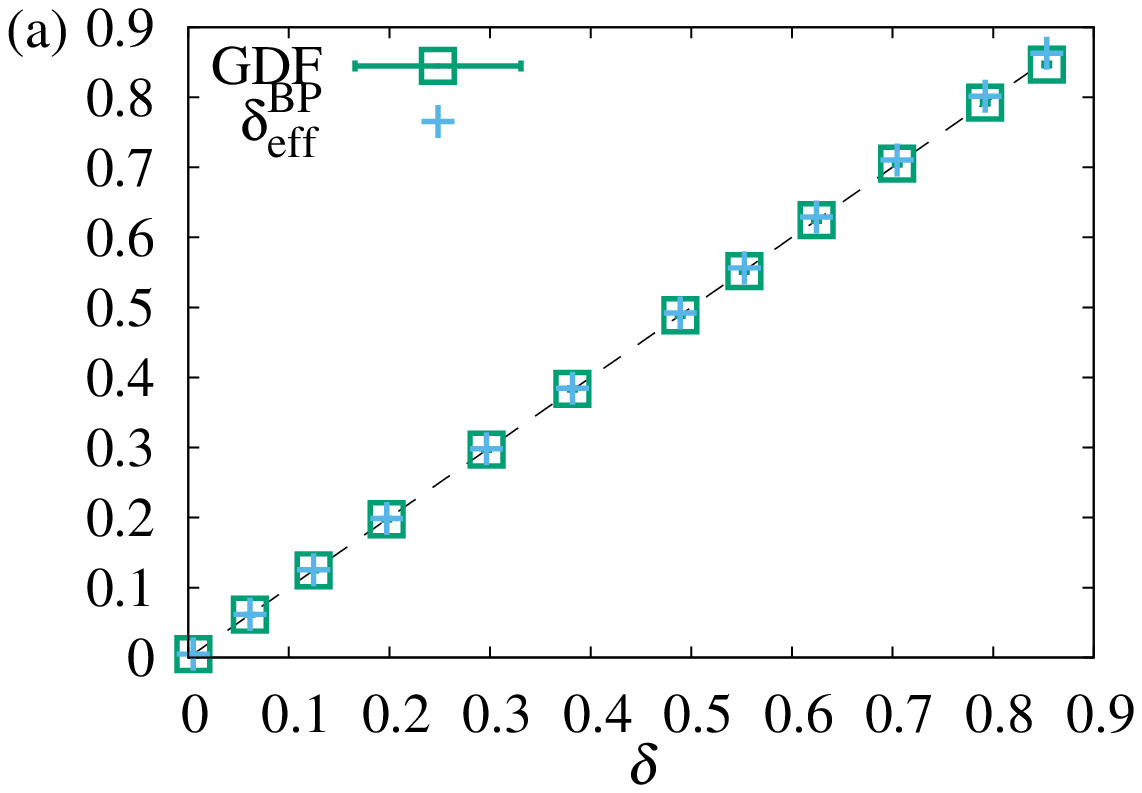}
\includegraphics[width=2in]{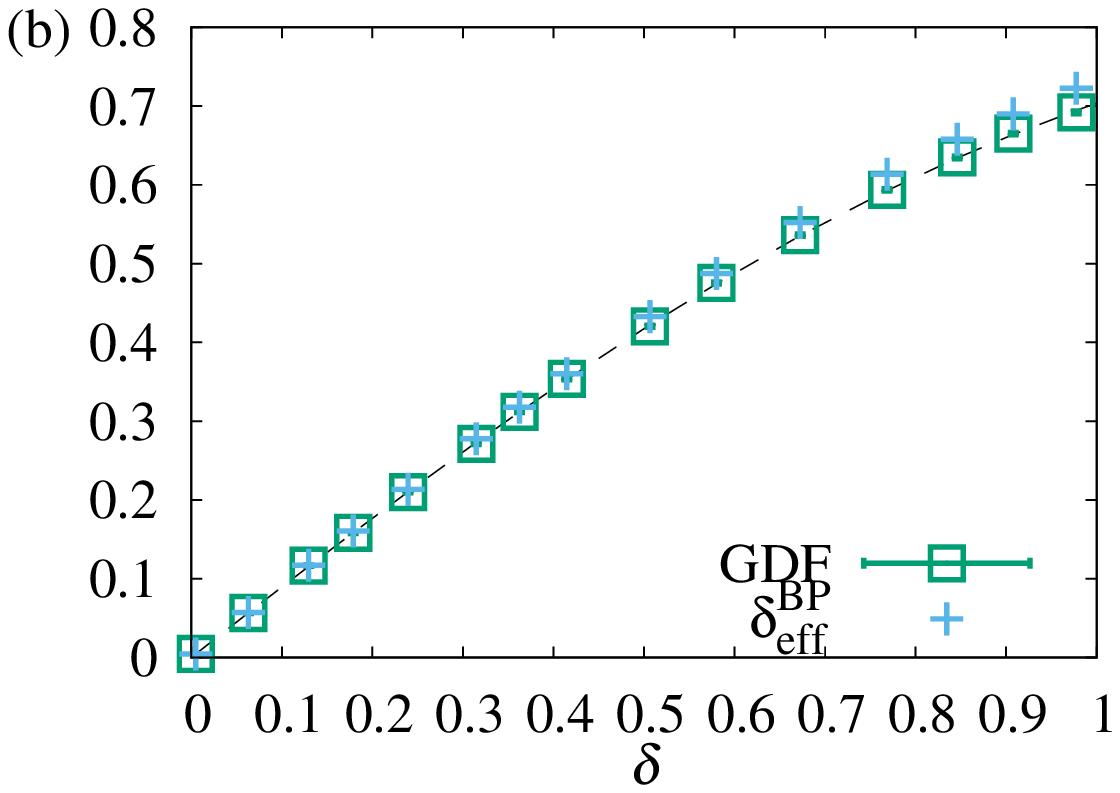}
\includegraphics[width=2in]{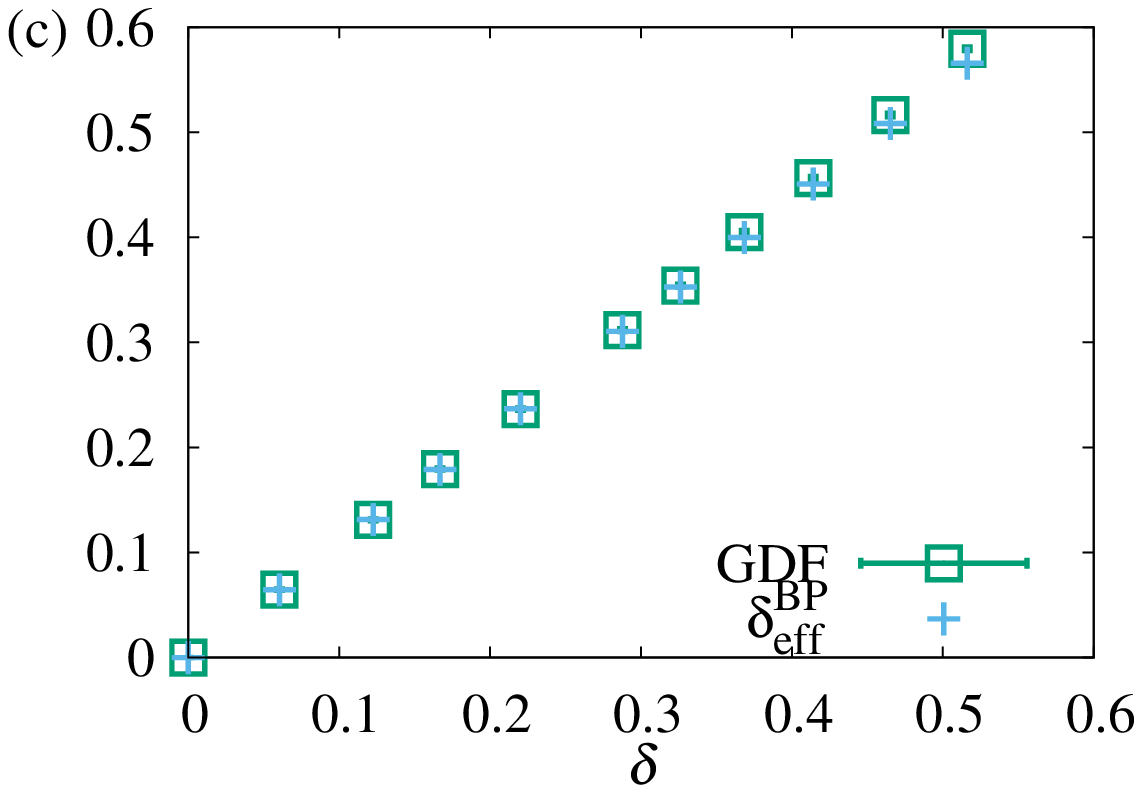}
\end{center}
\caption{$\delta$-dependence of GDF and $\delta_{\rm eff}^{\rm BP}$
at BP fixed point of $N=1000$, $M=500$ $(\alpha=0.5)$
for (a) $\ell_1$, (b) elastic net of $\eta_2=0.1$,
and (c) SCAD regularization of $a=0.5$ and $\lambda=1$
under the predictor matrix of example 1 with $c=0.5$.
We used 1000 samples of predictor matrices to calculate
the GDF and $\delta_{\rm eff}^{\rm BP}$ at the BP fixed point.
The $\delta$-region where the BP algorithm converges within $10^5$ steps
is shown. The dashed lines in (a) and (b) denote the results reported in \cite{L1_GDF}
and \cite{Zou_DThesis}.}
\label{fig:example1}
\begin{center}
\includegraphics[width=2in]{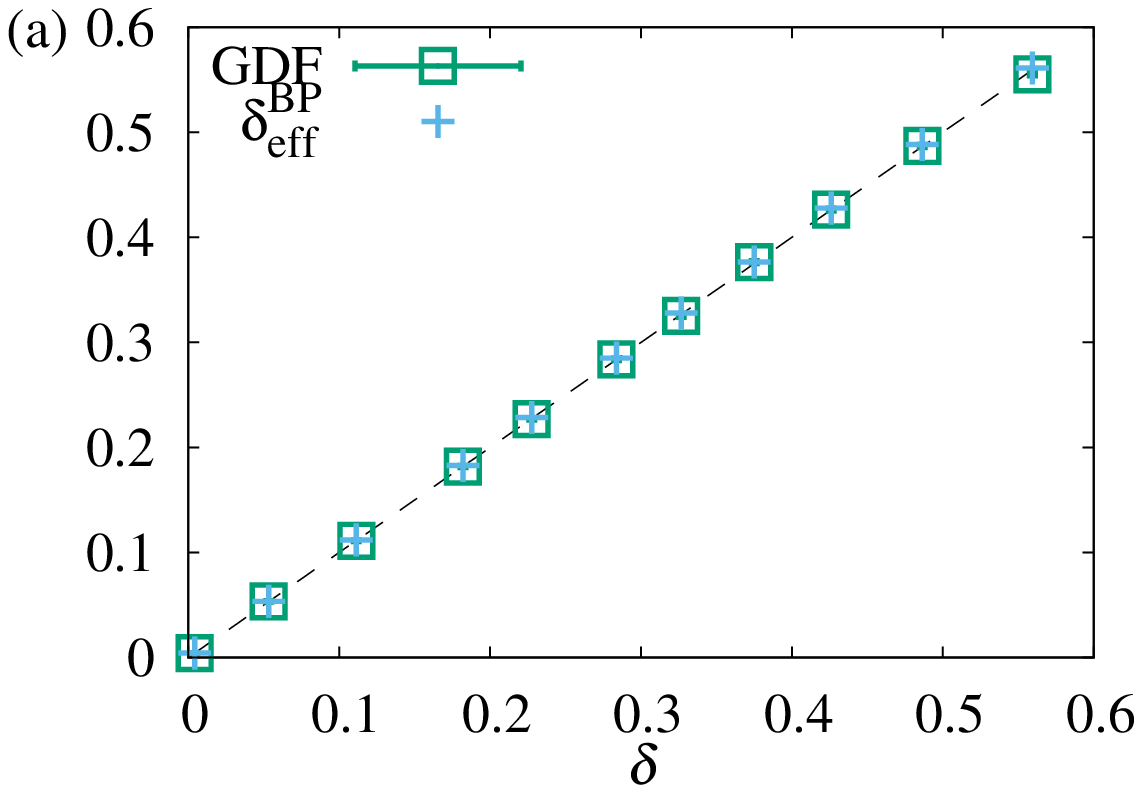}
\includegraphics[width=2in]{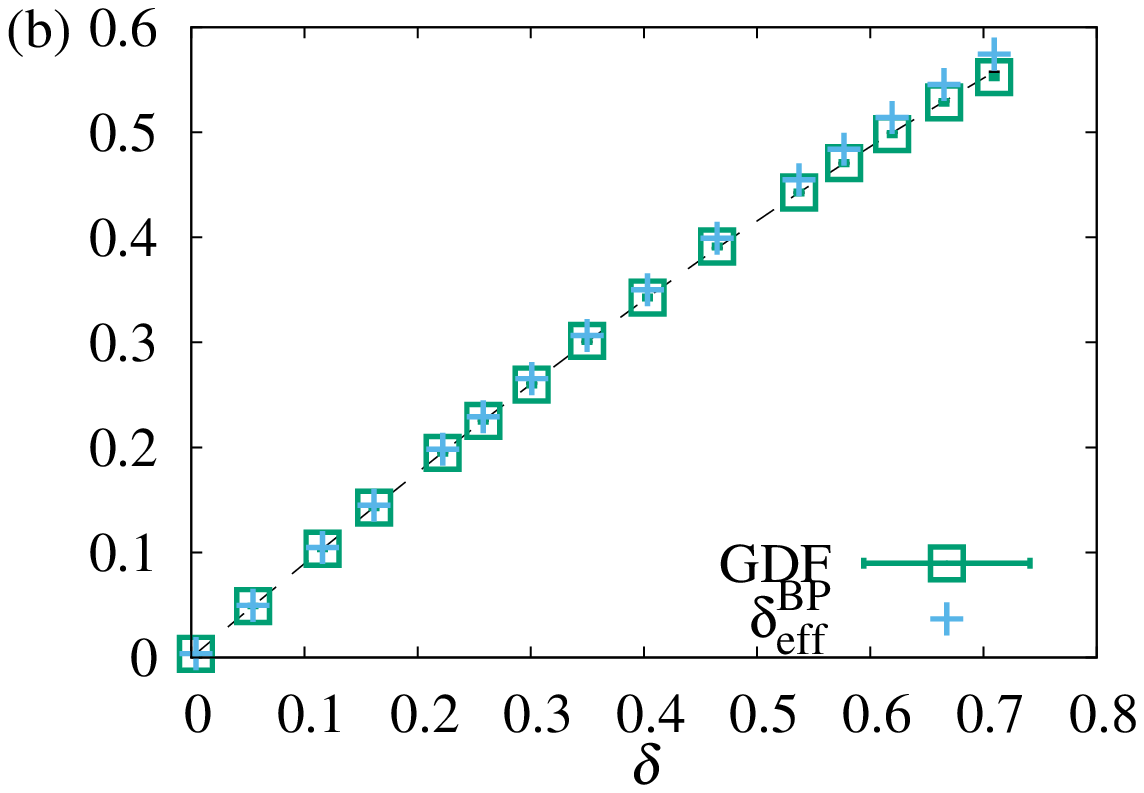}
\includegraphics[width=2in]{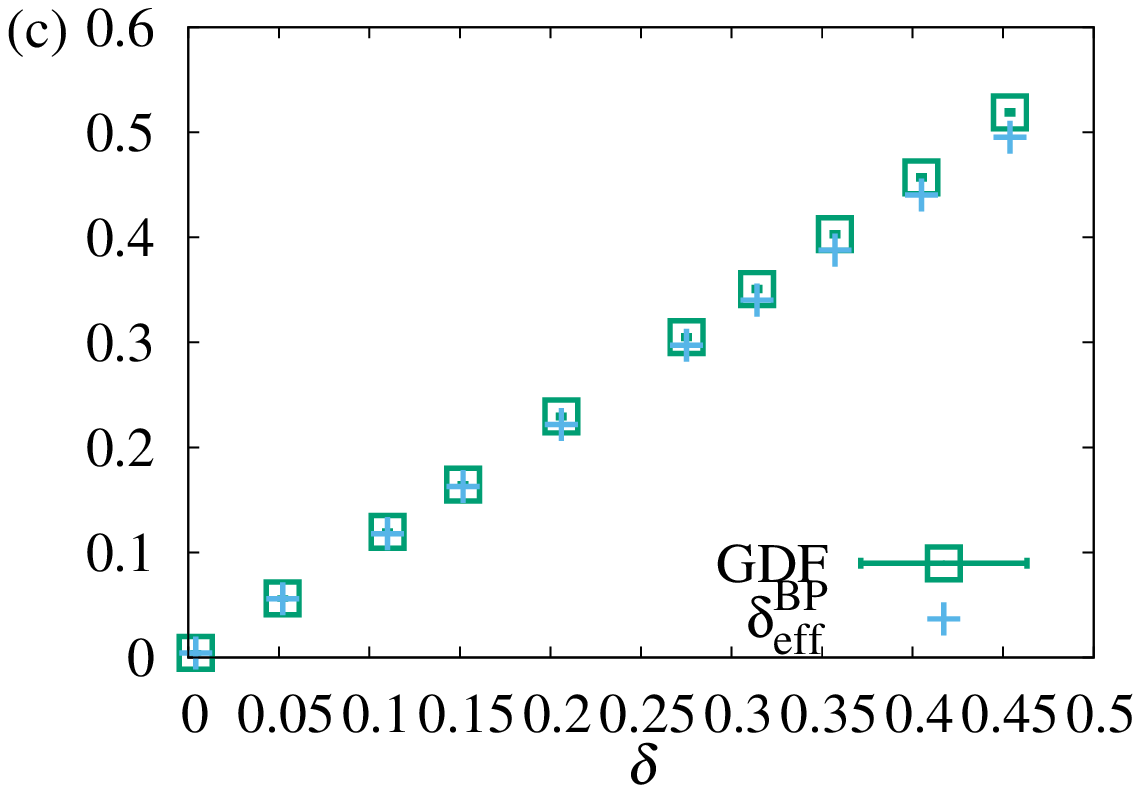}
\end{center}
\caption{$\delta$-dependence of GDF and $\delta_{\rm eff}^{\rm BP}$
at BP fixed point of $N=1000$, $M=500$ $(\alpha=0.5)$
for (a) $\ell_1$, (b) elastic net of $\eta_2=0.1$,
and (c) SCAD regularization of $a=0.5$ and $\lambda=1$
under the predictor matrix of example 2
with $T=125$.
We used 1000 samples of predictor matrices to calculate
the GDF and $\delta_{\rm eff}^{\rm BP}$ at the BP fixed point.
The $\delta$-region where the BP algorithm converges within $10^5$ steps
is shown. The dashed lines in (a) and (b) denote the results reported in \cite{L1_GDF}
and \cite{Zou_DThesis}.}
\label{fig:example2}
\end{figure}

Figures \ref{fig:example1} and \ref{fig:example2} 
show the $\delta$-dependence of GDF and $\delta_{\rm eff}^{\rm BP}$
at the BP fixed point for
$\ell_1$, elastic net, and SCAD regularization
at $N=1000$ and $\alpha=0.5$ $(M = 500)$.
The values of each point have been averaged over 1000 samples of
$\{\bm{y},\bm{A}\}$.
Under $\ell_1$ and elastic net regularization,
the GDF value calculated as the
expectation of the unbiased estimator
derived in \cite{L1_GDF,Zou_DThesis} is shown as a dashed line.
In both examples,
the correspondence between GDF and $\delta_{\rm eff}^{\rm BP}$
holds for each regularization,
although a small discrepancy appears due to finite-size effects at
large $\delta$.
Furthermore, the values of $\delta_{\rm eff}^{\rm BP}$ and GDF at the BP fixed point
are consistent with those of previous studies for $\ell_1$ and elastic net
regularization.
The parameters $c$ and $T$ in these examples do not
influence the results,
although they do affect the convergence of the BP algorithm.
These results 
imply that the correspondence between GDF and the effective fraction
of non-zero components
holds outside of Gaussian i.i.d. predictors.
For both examples, the convergence of the BP algorithm is
worse than with the Gaussian i.i.d. predictors,
particularly at large $\delta$.
Thus, the algorithm must be improved
to enable a discussion of the large-$\delta$ region
and application to other predictor matrices.

In the case of $\ell_0$ regularization,
the BP algorithm does not converge.
Thus, we cannot confirm the generality of the result
using the properties of BP fixed points.
The replica analysis for non-Gaussian i.i.d. predictor matrices
is a
necessary step towards 
verifying the generality of the result
for $\ell_0$ regularization.
Although the range of applicable predictor matrices for
replica analysis is narrower than that for the BP algorithm,
the analysis of rotationally invariant predictor matrices
offers a promising means
towards demonstrating the generality \cite{CS_Kabashima}.

\section{RS solution approximates the GDF for $\ell_0$ regularization}
\label{sec:l0_exhaustive}
For $\ell_0$ regularization, the RS solution is unstable in the whole parameter region,
but it is known that this solution generally 
approximates the true solution.
One can numerically obtain the exact solution of \eref{eq:lasso}
for $\ell_0$ regularization
by an exhaustive search, and
calculate the exact value of GDF
at small system sizes.
Comparing the GDF under RS analysis with its exact value,
we can evaluate the approximation performance of the RS solution.

\begin{figure}
  \begin{center}
    \includegraphics[width=3.3in]{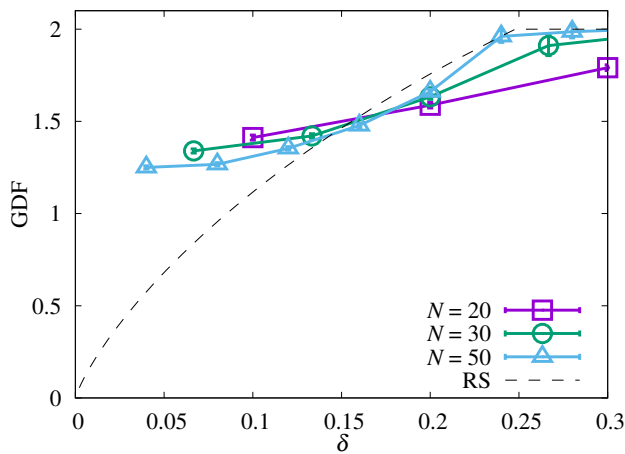}
  \end{center}
  \caption{Comparison between exact value of GDF at finite system sizes and GDF under RS analysis.}
  \label{fig:L0_exhaustive}
  \end{figure}

\Fref{fig:L0_exhaustive} 
compares the GDF approximated by the RS solution with its exact value
for $N=20,30$, and $50$
as calculated by 1000 samples of $\{\bm{y},\bm{A}\}$.
As $N$ increases, the exact GDF
approaches the RS solution,
although intense finite-size effects are observed
in the small-$\delta$ region.
For a comparison at larger system sizes,
we must develop a computationally feasible algorithm
for obtaining precise solutions of \eref{eq:lasso} for $\ell_0$
regularization,
but this is beyond the scope of the present paper.

\section{Summary and Conclusion}
\label{sec:summary}

We have derived the GDF
using a method based on statistical physics.
Within the range of the RS assumption,
the GDF is represented as $\chi\slash\hat{Q}^{-1}$,
where $\chi$ and $\hat{Q}^{-1}$
correspond to the rescaled variance around estimates
and
the variance of estimates
when the regularization term is omitted,
respectively.
This expression does not depend on the type of regularization, and indicates that GDF
can be regarded as the effective fraction
of non-zero components.

We applied our method for the derivation of GDF to $\ell_1$, elastic net, $\ell_0$, and SCAD regularization.
Our RS analysis was stable for $\ell_1$ and elastic net regularization
in the entire physical parameter region,
and the GDFs for these regularizations were consistent with
previous results.
This correspondence supports the validity of our RS analysis.
The model selection criterion of prediction error
was derived by combining the GDF with the training error.
Theoretical predictions in the RS phase were then algorithmically achieved
using the belief propagation method.

It has been implied that the equivalence between GDF and the ratio of the number of non-zero components
to the number of samples, $\delta$, only holds for $\ell_1$ regularization \cite{L1_GDF}.
Our representation of GDF as the effective
fraction of non-zero components
clarifies the
origin of
the additional component of the GDF
from the fraction of non-zero components.
\begin{itemize}
\item In $\ell_1$ regularization,
the GDF is given by $\delta$
because there is no factor that induces fluctuations
other than the non-zero components.

\item Elastic net regularization changes the variance of the components,
and so the GDF does not coincide with $\delta$.
However, as with $\ell_1$ regularization,
the non-zero components are the unique source of
fluctuations, and so
the correspondence between GDF and $\delta$
can be recovered by appropriately rescaling the estimates.

\item In $\ell_0$ regularization,
the discontinuity of the estimates
leads to additional fluctuations besides those caused by the non-zero components.
Hence, the GDF is greater than $\delta$.

\item In SCAD regularization,
the assignment of non-zero components to
the transient region between $\ell_1$-type estimates and
$\ell_0$-type estimates
induces additional components in the GDF.

\end{itemize}

For regularizations with AT instabilities in certain parameter regions (e.g. $\ell_0$, SCAD, and other non-convex regularizations),
it is generally necessary to construct the full-step RSB solution.
In the case of SCAD regularization,
model selection based on the prediction error
under RS analysis
can be achieved in the current problem setting.
Even when the RS solution is unstable, the prediction error gives a meaningful approximation
of the true value.

Further development of our method for the general function of prediction error \cite{Efron1986}
and real data will be useful for practical applications.
The BP algorithm discussed here can numerically calculate the 
GDF and model selection criterion for practical settings at reasonable computational cost.

\ack
The author would like to thank Yukito Iba, Yoshiyuki Kabashima, and Yoshiyuki Ninomiya
for insightful discussions and comments. This work was supported by JSPS KAKENHI No.25120013, 26880028 and 16K16131.

\end{document}